\documentclass[%
 reprint,
superscriptaddress,
 amsmath,amssymb,
 aps,
]{revtex4-1}

\usepackage{graphicx}
\usepackage{dcolumn}
\usepackage{bm}

\newcommand{\sigvec}{\vec{\sigma}}

\usepackage{braket}

\newcommand{\projector}[1]{\ket{#1}\!\!\bra{#1}}
\usepackage{comment}
\usepackage{hyperref}

\usepackage[color=green!40]{todonotes}
\setuptodonotes{inline}
\usepackage{multirow}
\begin{document}

\clearpage
\preprint{APS/123-QED}

\title{High-performance gates on trapped ion qubits\\ using counterpropagating pulse-shaped laser beams}

\author{Evangelos Piliouras}
 \affiliation{Department of Physics, Virginia Tech, Blacksburg, VA 24061, USA}
\affiliation{Virginia Tech Center for Quantum Information Science and Engineering, Blacksburg, VA 24061, USA}
\author{Hisham Amer}
 \affiliation{Department of Physics, Virginia Tech, Blacksburg, VA 24061, USA}
\affiliation{Virginia Tech Center for Quantum Information Science and Engineering, Blacksburg, VA 24061, USA}

\author{Susan M. Clark}
\author{Melissa C. Revelle}
\author{Edward C. Tortorici}
\affiliation{Sandia National Laboratories, Albuquerque, New Mexico 87123, USA}

\author{Matthew N. H. Chow}
\altaffiliation[Present address: ]{HRL Laboratory, LLC, Malibu, California 90265, USA}
\affiliation{Sandia National Laboratories, Albuquerque, New Mexico 87123, USA}
\affiliation{Department of Physics and Astronomy, University of New Mexico, Albuquerque, New Mexico 87131, USA}
\affiliation{Center for Quantum Information and Control, CQuIC, University of New Mexico, Albuquerque, New Mexico 87131, USA}

\author{Brandon Ruzic}
\author{Daniel S. Lobser}
\author{Brian K. McFarland}
\author{Christopher G. Yale}
\affiliation{Sandia National Laboratories, Albuquerque, New Mexico 87123, USA}

\author{Edwin Barnes}
 \affiliation{Department of Physics, Virginia Tech, Blacksburg, VA 24061, USA}
\affiliation{Virginia Tech Center for Quantum Information Science and Engineering, Blacksburg, VA 24061, USA}

\author{Sophia E. Economou}
 \affiliation{Department of Physics, Virginia Tech, Blacksburg, VA 24061, USA}
\affiliation{Virginia Tech Center for Quantum Information Science and Engineering, Blacksburg, VA 24061, USA}

\date{\today}

\begin{abstract}
Highly-localized light-matter interactions are necessary for scaling trapped-ion architectures. In hyperfine qubits, counterpropagating beams generate entangling gates by coupling with motion, but this effect is undesirable during single-qubit operations. For that reason, single-qubit gates are traditionally implemented with copropagating beams, and the coexistence of two beam geometries adds hardware and computational overhead. In an effort towards collective performance improvement with minimal overhead, we design and implement pulse-amplitude and dephasing robust dynamically corrected gates using Space Curve Quantum Control (SCQC) and compare them against the constant-amplitude gate implementation. We perform gate set tomography on a four-qubit trapped-ion register, and we discover more than 50\% error reduction when robust pulses are used. We find that counterpropagating robust gates often outperform their copropagating counterparts and reach error rates as low as $(3.59 \pm 1.25)\cdot 10^{-3}$, using diamond distance as a metric. This value establishes a laser-driven-gate error reference and is merely an order of magnitude higher than the best reported \textit{microwave} gate on a \textit{single} ion. Additional experiments reveal that robust pulses can effectively suppress non-Markovian errors that grow during runtime.
Our work challenges the widely accepted belief that copropagating gates should be preferred for their weak motional coupling and invites the adoption of high-performance robust pulses that suppress multiple noise sources of the trapped-ion error budget.
\end{abstract}
\maketitle
\section{Introduction}

Universal quantum computation requires a gateset of single- and two-qubit entangling gates \cite{NIELSENQuantumComputationQuantumInformation2010}. Across many platforms, implementing two-qubit gates necessitates the use of auxiliary states or modes outside the qubit subspace to mediate interactions, enabling switchable couplings and fast gate times. This is a common strategy in superconducting qubits, neutral atoms, and trapped-ion systems \cite{XUHighFidelityHighScalabilityTwoQubitGate2020, COLLODOImplementationConditionalPhaseGates2020,EVEREDHighfidelityParallelEntanglingGates2023, LEEPhaseControlTrappedIon2005}. However, while these auxiliary degrees of freedom are essential for entangling operations, they invariably introduce additional noise channels and error sources.

In trapped-ion platforms, collective motional modes of the ion chain serve as the standard mediator for two-qubit entangling gates \cite{SORENSENEntanglementQuantumComputationIons2000, SORENSENQuantumComputationIonsThermal1999}. Gate implementations are achieved through laser-driven Raman transitions, where counterpropagating beam geometries impart a net momentum kick to the ion crystal, coupling the qubit states to the collective motional modes and thereby enabling entanglement between qubits encoded in the internal states of the ions \cite{SORENSENEntanglementQuantumComputationIons2000, SORENSENQuantumComputationIonsThermal1999, YALERealizationCalibrationContinuouslyParameterized2025, CLARKEngineeringQuantumScientificComputing2021}.

For single-qubit gates, however, collective motional modes are not only unnecessary but also a source of errors. Managing these errors to achieve high-fidelity gates can incur a significant calibration and phase-stability overhead~\cite{LEEPhaseControlTrappedIon2005,DEBNATHDemonstrationSmallProgrammableQuantum2016,WRIGHTBenchmarking11qubitQuantumComputer2019,ZHUMultiroundQAOAAdvancedMixers2023}. Alternatively, single-qubit gates are often implemented in a way that avoids excitation of motional modes altogether. This is done either through microwave control or through copropagating laser configurations. While microwave control can reach impressively low single-qubit error rates \cite{SMITHSingleQubitGatesErrors102025}, localizing the mm-wave field at the micron scale of trapped ions remains a significant obstacle to scaling. For laser fields in the copropagating geometry, where the two Raman beams travel in the same direction, the wavevector difference between the ‘absorbed’ and ‘emitted’ photons vanishes, such that the net momentum transfer to the ion chain is negligible, ideally leaving the motional modes unperturbed \cite{WINELANDExperimentalIssuesCoherentQuantumstate1998, LEIBFRIEDQuantumDynamicsSingleTrapped2003}. Copropagating laser configurations thus avoid motional coupling but come with their own experimental overhead: the need for a modified setup compared to the counterpropagating one for two-qubit gates, which can in turn lead to its own phase-stability challenges~\cite{SHAFFERSampleefficientVerificationContinuouslyparameterizedQuantum2023,CHOWFirstorderCrosstalkMitigationParallel2024,YALERealizationCalibrationContinuouslyParameterized2025}. All three approaches to single-qubit gates (counterpropagating, copropagating, and microwave) thus add considerable complexity and hinder the scalability of trapped-ion platforms.

Here, we propose to address this challenge not by modifying the hardware or setup, but by treating the challenge as a control problem to be solved by pulse-shaping the lasers appropriately.  Rather than restricting the gate implementation to configurations that inherently avoid motional coupling, we instead allow counterpropagating pulses—which do generically couple to motional modes—and design pulse shapes that are intrinsically robust to the resulting motional noise. In this framework, the presence of the motional modes in the single-qubit context makes the Rabi rate phonon-number dependent, which amounts to a coherent amplitude error. The pulse can be engineered to be tolerant against amplitude noise and consequently suppress the effect of the motional modes. Our pulse-shaping strategy could also be applied to improve both copropagating-beam and microwave-driven single-qubit gates since calibration errors can give rise to coherent amplitude errors too.

In addition to motional noise in single-qubit gates implemented using counterpropagating pulses, implementing gates using lasers opens the door for more noise channels like Rabi rate fluctuations, differential AC Stark shifts, and beam-path instability \cite{MOUNTErrorCompensationSinglequbitGates2015, BROWNSinglequbitgateError1042011, GAEBLERHighFidelityUniversalGateSet2016,LEEPhaseControlTrappedIon2005}. Therefore, a pulse that is robust against multiple noise sources is necessary for high-fidelity gates implemented in either beam geometry. In order to achieve this goal of multi-noise source suppression, we use Space Curve Quantum Control (SCQC) \cite{BARNESDynamicallyCorrectedGatesGeometric2022,
BUTERAKOSGeometricalFormalismDynamicallyCorrected2021,
DONGDoublyGeometricQuantumControl2021,
LIDesigningArbitrarySingleaxisRotations2021,
NELSONDesigningDynamicallyCorrectedGates2023,
ZENGGeometricFormalismConstructingArbitrary2019,
ZHUANGNoiseresistantLandauZenerSweepsGeometrical2022, DAKISDynamicalErrorReshapingDualrail2025, ZENGGeneralSolutionInhomogeneousDephasing2018, ZENGFastestPulsesThatImplement2018,
WALELIGNDynamicallyCorrectedGatesSilicon2024,
AMERImplementingBenchmarkingDynamicallyCorrected2025}, and specifically its single-qubit gate design instantiation, the Bézier Ansatz for Robust Quantum (BARQ) control method that some of us previously developed \cite{PILIOURASAutomatedGeometricSpaceCurve2026}. In SCQC, the conditions necessary for noise suppression are expressed as global geometric properties of a space curve, while the control pulses are determined by that curve's local geometric quantities. BARQ provides a structured ansatz in which the target gate and dephasing-robustness are encoded directly in the curve's boundary conditions, while the curve's shape is optimized to achieve robustness against the remaining noise sources and for smooth pulse features. The result is a set of experimentally friendly dynamically corrected gates (DCGs) with guaranteed noise robustness \cite{KABYTAYEVRobustnessCompositePulsesTimedependent2014,
KHODJASTEHDynamicallyErrorCorrectedGatesUniversal2009,
KHODJASTEHDynamicalQuantumErrorCorrection2009,
KHODJASTEHAutomatedSynthesisDynamicallyCorrected2012,
KHODJASTEHArbitrarilyAccurateDynamicalControl2010,
BROWNArbitrarilyAccurateCompositePulse2004, POGGIUniversallyRobustQuantumControl2024}.

Our designs are validated experimentally on the QSCOUT trapped-ion testbed \cite{CLARKEngineeringQuantumScientificComputing2021}, where we implement detailed gate set tomography (GST) across a four-qubit register and demonstrate error rates (using a diamond distance metric) as low as $3.59 \cdot 10^{-3}$—representing more than a 3$\times$ reduction relative to the standard copropagating constant-envelope pulse. This value, to the best of our knowledge, establishes a diamond-distance error-rate record for laser-driven gates on trapped-ion architectures and is only one order of magnitude higher than the best reported single-qubit \textit{microwave} pulse error-rate \cite{BLUME-KOHOUTDemonstrationQubitOperationsRigorous2017}. These results demonstrate that robust counterpropagating pulses can match or exceed the performance of copropagating implementations while avoiding a significant calibration overhead, offering a pathway to reduce experimental complexity and improve the scalability of trapped-ion platforms.

The remainder of this paper is organized as follows. We begin with the system description, outline the robust pulse construction, and describe the minimal robust pulse calibration in Sec. \ref{experimental_considerations}. In Sec. \ref{gate_performance_comparison}, we summarize the results from GST and analyze the pulse performance in the aforementioned quantum runtime emulations. Finally, in Sec. \ref{conclusion}, we reflect on the acquired results and conclude this work.

\section{Experimental setup and noise-robust pulses}
\label{experimental_considerations}
We begin by briefly describing the experimental system and the noise model used to guide us in our robust pulse construction. We then outline the workflow for robust pulse generation and hardware implementation.

\subsection{System description and carrier transition noise model}
Our experimental setup is a ${}^{171}\text{Yb}^+$ four-ion chain hosted within the Quantum Scientific Computing Open User Testbed (QSCOUT) \cite{CLARKEngineeringQuantumScientificComputing2021} platform. The qubit is encoded in the hyperfine state manifold formed by the basis states $\ket{0} \equiv \ket{{}^2S_{1/2},F=0, m_F=0}$ and $\ket{1} \equiv \ket{{}^2S_{1/2},F=1, m_F=0}$ with an approximately 12.64 GHz energy splitting. Rabi oscillations are performed through Raman transitions with a 355~nm laser positioned about 33~THz away from the $\ket{0}$ and the ${}^2P_{1/2}$ transition. The calibration routine for constant pulses is described in Ref. \cite{YALERealizationCalibrationContinuouslyParameterized2025}.
Our gates are tested both under copropagating and counterpropagating beam configurations. The latter configuration is motion-sensitive and, as we show in Sec. \ref{momentum_kicks_section}, our robust pulses are resistant to errors stemming from this sensitivity. Before every experimental run, the ions are sideband cooled to reach $\bar n \approx 0.1$.

\begin{figure*}
    \centering
\includegraphics{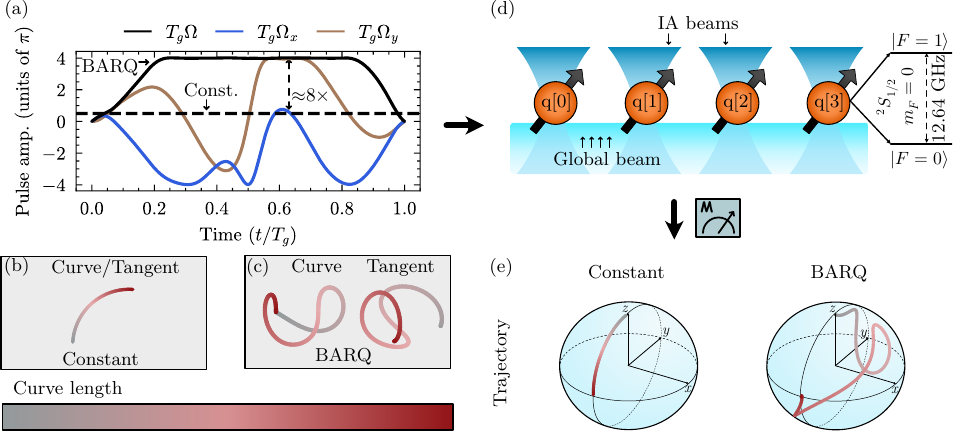}
    \caption{\textbf{Pulses, space curves, and experimental setup.} 
    (a) Control pulses with $\Omega_x = \Omega \cos\Phi$, $\Omega_y = \Omega \sin\Phi$. (b)--(c) Curve and tangent vectors for the constant pulse and the BARQ pulse, respectively. The hue indicates the time-arclength progression. The BARQ curve (c) is closed (dephasing robustness), by design, and its tangent curve traces zero area (amplitude error robustness), as can be verified visually from its approximate symmetry. The constant-amplitude $X_{\pi/2}$ pulse is represented as an arc (b) since constant amplitude translates to constant curvature. Lacking the above geometric characteristics, it is naturally susceptible to both types of error. (d) Experimental system. Our system is comprised of four ions that are controlled through both individual addressing (IA) beams and a global beam. This setup allows us to test both copropagating and counterpropagating gates, where the latter utilizes one IA beam and the global beam in opposite directions. The controls are fed to an RF-driven acousto-optic modulator that shapes the laser beams, enabling single-qubit rotations through Raman transitions executed in the $\ket{{}^2S_{1/2}, F=0, m_F=0}, \ket{{}^2S_{1/2}, F=1, m_F=0}$ hyperfine state manifold encoding the qubit. The BARQ pulse is appropriately scaled for the minimum possible gate time. (e) The Bloch sphere trajectories for a qubit initialized in its ground state reveal the subtleties of a robust pulse. The implementation of a robust pulse causes the spin vector to trace a lengthier path compared to the time-optimal constant-pulse rotation so that the robustness conditions are satisfied. The BARQ curve was optimized using the \texttt{qurveros} Python package \cite{PILIOURASQurverosSCQCBARQImplementation2025}. }
    \label{curve_system_desc}
\end{figure*}

Our system is described through the $L$-ion chain Hamiltonian \cite{WINELANDExperimentalIssuesCoherentQuantumstate1998}. 
To derive the system model, we proceed as follows. We first move to the rotating frame of the electronic and motional degrees of freedom. Considering, for simplicity, that only one ion is driven at a time (while the others remain idle), we apply the rotating wave approximation (RWA) on the interaction term. By assuming that the driving fields are resonant with the ion's electronic energy splitting, we obtain the following ion-motion interaction Hamiltonian (see Appendix \ref{ion_derivations}):
\begin{align}
    H_{\text{sys}} = \sum_{\vec n} \chi_{\vec n}(i \vec \eta) H_0(t) \otimes \projector{\vec n}.
\label{ion_chain_hamiltonian_rot}
\end{align}
We define $\vec n = [n_1 \, n_2 \dots n_{3L}]^T$ to be a vector that contains the occupation number for each motional mode and the associated state $\ket{\vec{n}} = \ket{n_1\,n_2\dots n_{3L}}$. The vector $\vec \eta$ contains the Lamb-Dicke (LD) parameters of each mode, and the function $\chi_{\vec n}(i \vec \eta)$ modulates the Rabi strength in each phonon subspace as
\begin{align}
    \chi_{\vec n}(i \vec \eta) &= \prod_{m=1}^{3L} \chi_{n_m}( \eta_m), \\
    \chi_{n_m}(\eta_m) &= \bra{n_{m}} 
    \exp[i\eta_{m}(a_{m} + a_{m}^\dagger)]
    \ket{n_{m}} .
\end{align}
When copropagating beams are utilized ($\vec \eta \approx \vec 0$), or the LD limit is assumed, where the spatial extent of the ion's motion is much smaller than the wavelength of the driving field (a case which also implies $||\vec \eta||_2 \ll 1$)\cite{WINELANDExperimentalIssuesCoherentQuantumstate1998}, each phonon subspace executes identical rotations under the Hamiltonian
\begin{align}
    H_0(t) = \frac{\Omega(t)}{2}[\cos\Phi(t)\sigma_x + \sin\Phi(t)\sigma_y],
    \label{qubit_hamiltonian}
\end{align}
where we define $\Omega(t)$ as the Rabi rate and $\Phi(t)$ as the phase of the drive. 

Depending on the driving scheme, $\Omega(t), \Phi(t)$ can either correspond to the amplitude and the phase of the microwave field or the effective quantities resulting from using lasers implementing Raman transitions (in the limit of large detuning away from a high-energy state outside the qubit subspace) \cite{LEIBFRIEDQuantumDynamicsSingleTrapped2003}.
The driving fields couple with the ion as $-\hat \epsilon \cdot \vec E(\vec r)  =\Omega \cos(\vec k \cdot \vec r-\omega_d t +\Phi(t)) (\sigma + \sigma^\dagger)$, where $\vec r$ is the ion's position, $\vec k$ is the $k$-vector of the field, and $\hat \epsilon$ denotes the field-ion coupling operator (e.g., a dipole operator). Here $\sigma, \sigma^\dagger$ are the atomic lowering and raising operators, respectively. In the microwave case, $\Omega(t)$ is the single-photon Rabi strength and $\Phi(t)$ is the phase of the field. The carrier transition is implemented by setting the driving frequency $\omega_d$ equal to the qubit energy splitting.
In the case of Raman transitions, we assume two lasers of the form $-\hat \epsilon_\pm \cdot \vec E_{\pm}(\vec r)  =\Omega_{\pm} \cos(\vec k_{\pm} \cdot \vec r-\omega_{\pm} t +\Phi_{\pm}(t)) (\sigma_{\pm,e} + \sigma_{\pm,e}^\dagger)$, where $\omega_{\pm}$ is detuned by $\Delta_e$ with respect to a higher-energy state $\ket e$. The field $\vec E_+$ drives the transition between the qubit ground state and $\ket{e}$, while $\vec E_-$ drives the transition between the qubit excited state and $\ket{e}$. The operator $\sigma_{\pm,e}$ de-excites the higher-energy state to either of the qubit states. For carrier transitions, we choose the Raman beat note ($\omega_+ - \omega_-$) equal to the qubit frequency, and in the large $\Delta_e$ limit the two-photon Rabi rate is $\Omega(t) = \Omega_+ \Omega_-/(2\Delta_e)$ and $\Phi(t)=\Phi_+(t)-\Phi_-(t)$. Our experimental setup supports both amplitude and phase modulation, for each ion,  through the drive of an acousto-optic modulator (AOM). 

In order to design robust controls, we start with a noise model that describes errors within the qubit subspace only. We assume the noise is quasi-static and modifies the qubit Hamiltonian by giving rise to an additive term $H_{\text{n}}$ such that the total qubit Hamiltonian becomes $H= H_0 + H_{\text{n}}$ where
\begin{align}
   H_{\text{n}} = \varepsilon H_0 + \frac{\delta_z}{2}\sigma_z.
\label{static_noise_hamiltonian}
\end{align}
This model captures shot-to-shot parameter fluctuations \cite{STECKMANNErrorMitigationShottoshotFluctuations2025} where the errors remain constant throughout the quantum evolution but change on a run-to-run basis. We assume additive frequency (dephasing) errors and multiplicative-type Rabi errors of the form $\Omega\to(1+\varepsilon) \Omega$. The former noise source stems mostly from potentially uncompensated AC Stark shifts present in Raman transitions or stray magnetic fields. Additionally, slow phase variations arising from beam-path instabilities can produce a similar error term \cite{LEEPhaseControlTrappedIon2005} in the appropriate frame. The multiplicative-type Rabi rate errors are usually a byproduct of amplitude calibration errors, where statistical uncertainty is unavoidable. Beam spatial distortion and potential spill-over during simultaneous qubit drive \cite{CHOWFirstorderCrosstalkMitigationParallel2024} can cause effects that can be captured by the same noise term.

\subsection{Robust pulse construction}
\subsubsection{Noise influence and space curves}
In order to construct pulses that dynamically suppress errors during the quantum evolution, we begin by breaking the problem into two parts. Considering that the ideal evolution $U_0$ is generated by the controls of Eq.~\eqref{qubit_hamiltonian}, $i\dot{U}_0 = H_0U_0$, we decompose the non-ideal evolution $U$ generated by $H=H_0 +H_{\text{n}}$ as $U = U_0U_I$, where $U_I$ captures the effects of noise. In this way, we separate the influence of noise from the task of designing a gate under ideal controls.  Using the average Hamiltonian theory \cite{BRINKMANNIntroductionAverageHamiltonianTheory2016}, noise-suppressing controls must satisfy
\begin{align}
    \int_0^{T_g} dt \, U_0^\dagger H_{\text{n}} U_0 =0 \iff \nonumber \\ \int_0^{T_g} dt \left[ \varepsilon \, U_0^\dagger H_0 U_0 +\frac{\delta_z}{2 }U_0^\dagger\sigma_zU_0 \right]=0.
    \label{noise_suppress_cond}
\end{align}

Many different waveforms can satisfy the above equation while achieving the desired operation $U_g$ in the ideal case ($U_0(T_g)=U_g$). In this work, we utilize the Space Curve Quantum Control (SCQC) formalism \cite{BARNESDynamicallyCorrectedGatesGeometric2022}, a framework that allows us to characterize control robustness as global geometric properties of space curves. In SCQC, we define a \textit{space curve} $\vec r(t)$ using the three time-dependent coefficients that arise when we expand the following integral in a Pauli basis:
\begin{align}
    \int_0^{t} dt' \, U_0^\dagger \sigma_z U_0 = [\vec r (t) -\vec r(0)] \cdot \sigvec,
    \label{scqc_main_eq}
\end{align}
where $\sigvec = \begin{bmatrix}
    \sigma_x & \sigma_y & \sigma_z
\end{bmatrix}^T$ is the Pauli vector. We can then rewrite Eq.~\eqref{noise_suppress_cond} as \cite{NELSONDesigningDynamicallyCorrectedGates2023}
\begin{align}\label{eq:noise_cancellation}
    \varepsilon \int_0^{T_g} dt\, \vec T \times \dot{\vec T} +  \frac{\delta_z}{2}\left[\vec{r}(T_g)-\vec r(0)\right] = \vec 0,
\end{align}
where we defined the tangent vector of the space curve $\vec r$  as $\vec T \equiv \dot {\vec r}$. This equation indicates that evolution time is equivalent to the arc length along the curve. The quantum operation $U_0(T_g)$ is determined primarily by the relative orientation of the initial- and final-time tangent vectors, namely the quantity $\vec T(0)\cdot \vec T(T_g)$. More details are given in Appendix \ref{scqc_barq_review}.

It is immediately evident from Eq.~\eqref{eq:noise_cancellation} that the dephasing error ($\delta_z$) term is suppressed when the space curve is closed ($\vec{r}(T_g)= \vec{r}(0)$). On the other hand, the first term in Eq.~\eqref{eq:noise_cancellation} (stemming from the Rabi rate error $\varepsilon$) is a 3-component vector, $\int_0^{T_g} dt\, \vec T \times \dot{\vec T}=A_x\hat x+A_y\hat y+A_z\hat z$, where, e.g., $A_x$ is the two-dimensional area enclosed by the tangent curve when it is projected onto the plane orthogonal to $\hat x$. When $A_x$, $A_y$, $A_z$ all vanish, amplitude robustness is achieved \cite{NELSONDesigningDynamicallyCorrectedGates2023}. Once a space curve satisfying these geometric properties is constructed, we compute the curve's curvature $\kappa(t)$ and torsion $\tau(t)$ to obtain the control fields that implement the corresponding robust evolution using the following relations:
\begin{align}
    \Omega(t) &= \kappa(t) = ||\dot{\vec T}||_2, \\
    \dot \Phi(t)&=\tau(t)= (\vec T \times \dot {\vec T})\cdot \ddot {\vec T}/\kappa^2(t).
\end{align}
Next, we discuss how to systematically construct space curves with the desired properties.

\subsubsection{Automated space curve design}
To facilitate the design of suitable space curves, we use the Bézier Ansatz for Robust Quantum (BARQ) control method \cite{PILIOURASAutomatedGeometricSpaceCurve2026}. BARQ utilizes a particular space curve ansatz, called the Bézier curve, which is expressed as
\begin{align}
    \vec r(x(t)) = \sum_{j=0}^{n} \vec w_j g_{j,n}(x(t)),\qquad x \in [0,1].
\end{align}
The curve's shape is dictated by the position of the control points $\vec w_j$, which serve as points of attraction as the space curve is traversed from $x=0$ to $x=1$, and the $g_{j,n}(x(t))$ are polynomial basis functions.
Once a space curve is constructed, the function $x(t)$ is determined by enforcing that the time variable $t$ is the arc length parameter of the curve, as required in SCQC.

The motivation behind using BARQ is that we can enforce some constraints at the outset by fixing a subset of the control points, thus simplifying the design process. In fact, both the target operation and the dephasing robustness (closed-curve condition) are imposed by appropriately distributing some of the control points. The rest are freely optimized for robustness against amplitude errors while minimizing the maximum pulse amplitude. The optimization is done using the \texttt{qurveros} Python package \cite{PILIOURASQurverosSCQCBARQImplementation2025}, and more details are provided in Appendix \ref{scqc_barq_review}.

The resulting curve and controls are shown in Fig.~\ref{curve_system_desc}. In Fig.~\ref{curve_system_desc}a, we show the SCQC-generated control fields that implement the noise-robust evolution. The resulting $X_{\pi/2}$ pulse is smooth but has a larger (time-normalized) maximum pulse amplitude. In Fig.~\ref{curve_system_desc}c, the BARQ-optimized controls are associated with a closed curve which guarantees dephasing robustness, while its tangent's approximate symmetry signifies amplitude robustness (the total traced area vanishes). In stark contrast, the constant pulse is associated with a circular arc for both the curve and its tangent. Since the required geometric features are absent, neither desired robustness characteristic is achieved. Another perspective in understanding the robust evolution is depicted in the Bloch sphere trajectory in Fig.~\ref{curve_system_desc}e. While the constant pulse reaches the final state by taking the time-optimal route, the robust pulse requires a more elaborate path so that both robustness conditions are satisfied.

\subsection{Calibrating robust pulses for quantum hardware}
\label{calibration_method}
In the absence of experimental non-idealities, the robust quantum evolution can be realized at any timescale. We can simply rescale the SCQC control Hamiltonian and match it with the quantum device under test (QDUT) Hamiltonian:
\begin{align}
    T_g^{\text{QDUT}} H_0^{{\text{QDUT}}} &= T_g^{\text{SCQC}} H_0^{\text{SCQC}} \nonumber \iff \\ 
T_g^{\text{QDUT}} \Omega^{\text{QDUT}} &= T_g^{\text{SCQC}}\Omega^{\text{SCQC}}.
\label{equivalent_hamiltonians}
\end{align}
We shall assume that the QDUT is calibrated, and without loss of generality, we can obtain access to the parameters that realize a $\pi$-pulse. Expressed mathematically, we have $\Omega_\pi^{\text{QDUT}} T_\pi^{\text{QDUT}} = \pi$, where $\Omega_\pi^{\text{QDUT}}, T_\pi^{\text{QDUT}}$ are the associated Rabi rate and gate time. The procedure can be modified for an arbitrary angle of rotation and pulse shape.

In order to transfer the robust pulse to hardware, one can tune the control apparatus based on either a reference Rabi strength or a given gate duration. For this work, we use an amplitude scale parameter $s_\Omega$, whose role is to quantify the ratio of the maximum pulse amplitude to that obtained from the calibration data: 
\begin{align}
    \max\{|\Omega^{\text{QDUT}|}\} = s_\Omega\, \Omega_\pi^{\text{QDUT}}.
    \label{amp_scale_eq}
\end{align}
The chosen amplitude scale will naturally determine the experimental gate time. To complete the transfer of the robust pulse to hardware, we apply Eq. \eqref{amp_scale_eq} to Eq. \eqref{equivalent_hamiltonians} and obtain 
\begin{align}
    \Omega^{\text{QDUT}} &= \Omega_\pi^{\text{QDUT}} \times s_\Omega \times \frac{T_g^{\text{SCQC}}\Omega^{\text{SCQC}}}{\max \{ T_g^{\text{SCQC}}\Omega^{\text{SCQC}}\}}, \\
T_g^{\text{QDUT}} &= T_\pi^{\text{QDUT}} \times s_\Omega^{-1} \times \max \{ T_g^{\text{SCQC}}\Omega^{\text{SCQC}}\} /\pi.\label{eq:gate_time}
\end{align}
For a fixed target operation, an increased Rabi strength will reduce the required gate time. The robust pulse's gate time depends on  $s_\Omega$ and $\max \{ T_g^{\text{SCQC}}\Omega^{\text{SCQC}}\}$. When compared to the reference pulse's gate duration, the value of $\max \{ T_g^{\text{SCQC}}\Omega^{\text{SCQC}}\}$  dictates how much \textit{slower} the robust gate will be. This fact is already hinted at by the lengthier Bloch sphere trajectory in Fig. \ref{curve_system_desc}e. 

From Eq.~\eqref{eq:gate_time}, we see that $s_\Omega$ can be viewed as a parameter that controls the gate time, such that larger values of $s_\Omega$ lead to faster gates. We need to choose the highest possible value for $s_\Omega$ before other effects, not captured by the employed model, start to degrade the quantum gate quality. Choosing $s_{\Omega}=1$ might be a rather limiting option since executing the robust pulse while retaining the same maximum pulse amplitude will result in an approximately 8$\times$ slower gate (see Fig. \ref{curve_system_desc}a).
\begin{figure}
    \centering
\includegraphics{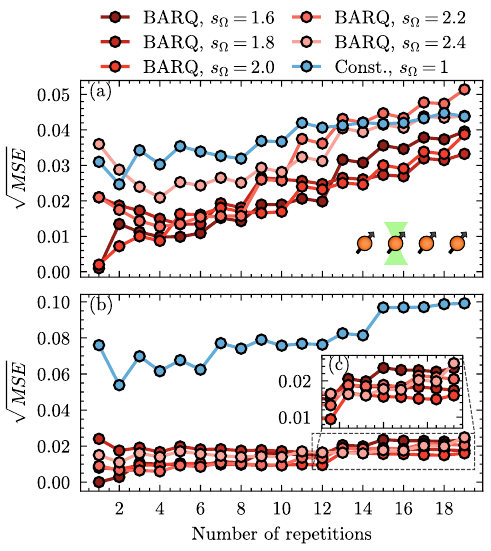}    
\caption{\textbf{Error accumulation after repeated applications of $X_{\pi/2}$ gates for various amplitude scales $s_{\Omega}$.} To choose the amplitude scale of our noise-robust pulses, the same $X_{\pi/2}$ gate is repeated up to $K=19$ times, only on qubit q[1], which is initialized in $\ket{0}$, and the excited state probability is measured after each number $k$ of repetitions. The mean-squared error ($MSE$) is calculated as the average squared difference between the measured values and the expected (nominal) probability values for every $k\le K$. (a) Experimental results for the copropagating configuration. When $s_\Omega > 2$, errors accumulate slightly more for the robust BARQ pulses (red) compared to the non-robust constant-pulse case (blue), which we attribute to pulse distortions from the AOM. This trend disappears at lower values of $s_\Omega$, and with the robust pulse designed to suppress two noise types, it accumulates a smaller $MSE$. (b) Experimental results for the counterpropagating configuration and a zoomed view (c) for the last 8 repetitions. The robust pulse outperforms the constant pulse in all amplitude scales with a significant error difference. In this configuration, the dominant source of error is the ion-motion coupling, which contributes to amplitude errors that remain substantial in the constant-pulse case.}  
    \label{calibration_mse}
\end{figure}

\begin{table}
    \begin{ruledtabular}
    \begin{minipage}{0.75\columnwidth}
        \begin{tabular}{lcc}
 Beams& \multicolumn{2}{c}{Gate time (\textmu s)} \\
 & Constant & BARQ \\
 &  &  \\
Coprop. & 12.5 & 50 \\
Counterprop. & 5 & 20 \\
\end{tabular}
    \end{minipage}
    \caption{Gate duration summary for the $X_{\pi/2}$ gates.}
    \label{gate_time_summary}
\end{ruledtabular}
\end{table}

In order to choose the value of $s_{\Omega}$, we follow a procedure similar to the one employed in Ref. \cite{SMITHSingleQubitGatesErrors102025} and apply our gate $U$ up to $K$ times (with $K=19$, as we find that this is already sufficiently large for gate errors to become appreciable), measuring the overlap with the excited state, $|\bra{1}U^k\ket{0}|^2$, after the $k$-th application, $k=1 \dots K$. This simple experiment reveals the rough accumulation of errors and guides us towards the optimal value for $s_\Omega$. While this calibration can be done for every qubit, we chose to execute it for simplicity only on qubit q[1].
The gate is written as $U(T_g^{\text{QDUT}})=U_g U_I(T_g^{\text{QDUT}})$ and in the experimental setting, each value of $s_\Omega$ will lead to a different $U_I(T_g^{\text{QDUT}})$. To quantify the accumulated errors, we calculate the running mean squared error (MSE) with respect to the nominal probability values. For every sequence length, $k=1 \dots K$, we have
\begin{align}
    MSE[k]  = \frac{1}{k}\sum_{k'=1}^k (|\bra{1}U^{k'}\ket{0}|^2 - p_{k'})^2,
\end{align}
where $p_{k'}=|\bra{1}U_g^{k'}\ket{0}|^2$ is the probability after $k'$ applications of the ideal $U_g$.

Here, we focus on the case where $U_g$ is an $X_{\pi/2}$ gate, in which case the nominal probabilities are $p_{k'} = \{ 0.5, 1,0.5,0\}$ for $k'\,  \text{mod} \, 4=0,1,2,3$, respectively. The results for this gate are summarized in Fig.~\ref{calibration_mse}, where the error accumulation in both copropagating (Fig.~\ref{calibration_mse}a)  and counterpropagating (Fig.~\ref{calibration_mse}b) configurations is shown. In the copropagating beam geometry, all pulses exhibit similar behavior, with the BARQ pulse performing slightly better for $s_\Omega \le2$. Larger amplitudes can yield laser strengths that induce AOM pulse distortions \cite{YALERealizationCalibrationContinuouslyParameterized2025}. We believe this is the main reason for the gradual error accumulation when $s_\Omega>2$. The picture for the counterpropagating beams is different. The constant pulse accumulates roughly four times as much error as the BARQ pulse does, for all amplitude scales. This behavior is expected since the ion-motion coupling reduces the average driving Rabi strength. In contrast, the BARQ pulses are robust to motion-induced amplitude errors (see Sec. \ref{momentum_kicks_section}), which explains why the $MSE$ values in both beam setups are close to each other.

From the above observations, we choose to run our pulses with $s_\Omega=2$ for both beam geometries. While it might be more prudent to find the best $s_\Omega$ for every qubit in every beam configuration, we instead opt for a simple, one-time calibration. This allows us to investigate error suppression with variable qubit calibration parameters. The gate durations are summarized in Table \ref{gate_time_summary}, rounded to the closest integer.

In closing this section, it is important to note that BARQ can design noise-robust gates of arbitrary angle and axis of rotation. The $X_{\pi/2}$ pulse serves as a simple demonstration since it is a building block for single-qubit gates. While cascading two robust $X_{\pi/2}$ pulses can synthesize a robust $\pi$-pulse, one risks unnecessarily prolonging the required gate time. For that purpose, a robust $\pi$-pulse should be designed directly so that $T_g \Omega_{\max}$ can be made smaller. An example BARQ $\pi$-pulse with the same robustness characteristics is found in Ref. \cite{PILIOURASAutomatedGeometricSpaceCurve2026}. Whether or not a single-shot robust gate is preferred over the composition of two robust gates is a matter of how each gate shapes the noise influence ($U_I(T_g)$) and other effects that manifest from the choice of $s_\Omega$.

\section{Gate performance comparison}
\label{gate_performance_comparison}
In the experiments testing and comparing gate performance described below, we calibrated only the constant-amplitude pulse. For the BARQ pulses, we set $s_\Omega = 2$ as discussed in the previous section. We performed two classes of experiments within an approximately five-month window. The first class characterized all four qubits using gate set tomography (GST), while the second class studied the gate behavior under realistic quantum computer runtime conditions. The choice of $s_\Omega = 2$ was made based only on the results of q[1] taken at the beginning of the data acquisition period and remained fixed for the remainder of the experiments. All qubits showed performance benefits despite the large data-acquisition time window. Using data taken from a single qubit to calibrate the BARQ pulses for all qubits is very efficient. Moreover, we believe that the simple gate repetition experiment we employed (and described above) is enough to indicate the best value of $s_\Omega$ compared to more elaborate experiments (e.g., tomography).

\subsection{Gate set tomography}
\label{gst_results_sec}

We saw above that for most choices of $s_\Omega$, the robust pulses maintain a smaller accumulated error (see Sec. \ref{calibration_method}).
In order to obtain more concrete results, we implement a gate set tomography (GST) experiment \cite{NIELSENGateSetTomography2021, BLUME-KOHOUTDemonstrationQubitOperationsRigorous2017} to extract various gate error metrics. We assume that the gate is implemented imperfectly, and a noise channel $\mathcal{E}$ acts before or after the ideal operation. The average gate/channel infidelity is defined as \cite{WALLMANNoiseTailoringScalableQuantum2016}:
\begin{align}
    r(\mathcal{E}) = 1 - \int d\rho \, \text{tr}(\mathcal{E}(\rho)\rho),
\end{align}
where a uniform averaging is performed over pure states $\rho$. For fault-tolerant quantum error correction, average gate infidelity can be an inadequate metric because logical failure probabilities depend on the particular input states, syndrome measurements, and error-propagation mechanisms involved, rather than on a state- and measurement-averaged gate error alone~\cite{SANDERSBoundingQuantumGateError2015}. For these reasons, the diamond distance from the identity is preferred, especially when the noise channel includes both stochastic and coherent errors \cite{BLUME-KOHOUTDemonstrationQubitOperationsRigorous2017, AHARONOVFaultTolerantQuantumComputationConstant2008}. The diamond distance from the identity is defined as \cite{WALLMANNoiseTailoringScalableQuantum2016}
\begin{align}
    \epsilon_{\diamond}(\mathcal{E}) = \frac{1}{2}||\mathcal{E} - \mathcal{I}||_{\diamond}= \sup_\rho  \frac{1}{2} ||(\mathcal{E}\otimes I_d- I_{d^2})(\rho)||_1,
\end{align}
where $d$ is the system's Hilbert space dimension, and $||A||_1 = \text{tr}(\sqrt{A^\dagger A})$. Here $\rho$ is a quantum state that lives in a Hilbert space of dimension $d^2$. The diamond distance is not experimentally measurable, while the most accessible performance metric is the average gate infidelity (with some caveats depending on the method used \cite{PROCTORWhatRandomizedBenchmarkingActually2017}). We can relate these two quantities by \cite{WALLMANNoiseTailoringScalableQuantum2016,WALLMANRandomizedBenchmarkingConfidence2014}
\begin{align}
    r(\mathcal{E}) \frac{d+1}{d} \le \epsilon_{\diamond}(\mathcal{E}) \le \sqrt{d(d+1)}\sqrt{r(\mathcal{E})}.
\end{align}
It is clear from these relations that an optimistic average gate infidelity estimate does not necessarily translate to low error rates. In fact, the two metrics become linearly proportional only when the noise channel introduces purely incoherent noise \cite{BLUME-KOHOUTDemonstrationQubitOperationsRigorous2017, SANDERSBoundingQuantumGateError2015}. But trapped-ion systems can also contain coherent errors, in which case the two error metrics will differ.

\begin{figure}
    \centering
    \includegraphics{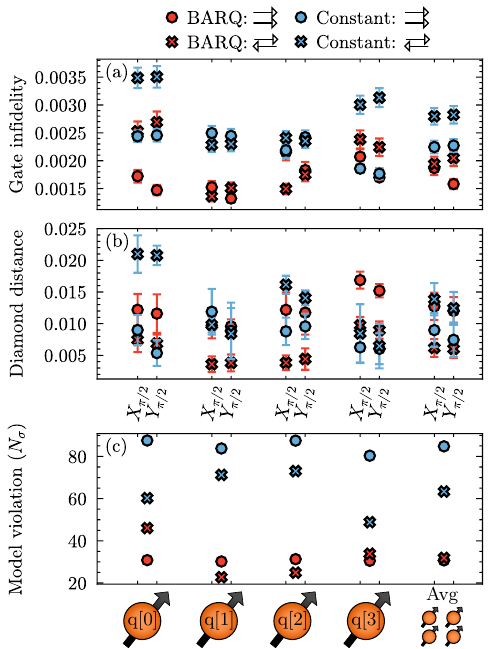}
    \caption{\textbf{Gate set tomography results.} Results for each qubit are shown along the horizontal axis, and the average over all four is shown on the right side of each panel. The error bars correspond to 95\% confidence intervals. (a) Gate infidelities and (b) diamond distances for each gate in the gateset (except the virtual $Z_{\pi/2}$) are shown. For all qubits except q[3], the robust pulses (red) exhibit lower gate infidelities compared to constant pulses (blue). The diamond distances follow a similar pattern, with nearly 3$\times$ improvements for some qubits. (c) GST fit model violation. The BARQ pulse can effectively suppress non-Markovian errors given the roughly 2$\times$ reduction of $N_\sigma$. Remarkably, counterpropagating robust pulses often perform better than both their copropagating counterparts and the constant-pulse implementations.}
    \label{gst_results}
\end{figure}

\begin{table*}
   \begin{ruledtabular}
   \setlength{\tabcolsep}{10pt}
    \begin{tabular}{llllllllllll}
 &  & \multicolumn{5}{c}{Diamond distance (BARQ/Constant)} & \multicolumn{5}{c}{Infidelity (BARQ/Constant)} \\
 & Qubit & q[0] & q[1] & q[2] & q[3] & Avg & q[0] & q[1] & q[2] & q[3] & Avg \\
Gate & Beams &  &  &  &  &  &  &  &  &  &  \\
\multirow[l]{2}{*}{$X_{\pi/2}$} & Coprop. & 1.35 & 0.82 & 1.39 & 2.69 & 1.42 & 0.71 & 0.61 & 0.98 & 1.11 & 0.83 \\
 & Counterprop. & 0.35 & 0.37 & 0.24 & 1.15 & 0.44 & 0.72 & 0.60 & 0.62 & 0.79 & 0.69 \\
\multirow[l]{2}{*}{$Y_{\pi/2}$} & Coprop. & 2.17 & 1.08 & 1.23 & 2.52 & 1.61 & 0.60 & 0.54 & 0.76 & 0.96 & 0.70 \\
 & Counterprop. & 0.33 & 0.45 & 0.31 & 1.39 & 0.48 & 0.77 & 0.66 & 0.74 & 0.72 & 0.73 \\
\end{tabular}

\caption{\textbf{Diamond distance and infidelity ratios for robust BARQ versus non-robust constant pulses.} The left side of the table shows the ratio of the BARQ diamond distance over the constant-pulse diamond distance obtained from GST for each gate, beam geometry, and qubit, as well as the results averaged over all 4 qubits. Similarly, the right side of the table shows ratios of infidelities for the BARQ and constant pulses. In all cases, these ratios are obtained from the data shown in Fig.~\ref{gst_results}.}
    \label{gst_data_ratios}
    \end{ruledtabular}
\end{table*}

GST is a comprehensive characterization tool, but it can become experimentally expensive as the number of qubits grows. Since we aim to characterize single-qubit rotations, the GST's rigor will give us strong evidence regarding the benefits of adopting robust pulses. We perform extended linear GST (eLGST) on all four qubits, driving one qubit at a time while the rest remain idle. The presence of the other qubits contributes to motional spectator modes that can affect the average Rabi strength. In order to obtain an adequate margin to estimate gate-performance metrics, we choose a gate sequence length of 2048 with 200 shots. This gives an error in estimating the gates on the order of $10^{-5}$ \cite{NIELSENGateSetTomography2021}.
Our gate set is defined as
 \begin{align}
     \mathcal{G} = \{\rho_0 = \projector{0};X_{\pi/2}, Y_{\pi/2}, Z_{\pi/2}; \nonumber \\ 
     E_0 = \projector{0}, E_1 = \projector{1} \},
 \end{align}
where $\rho_0$ is the initial state, and $E_{0/1}$ are measurement operators in the $Z$-basis. The same set holds for all qubits. The $X_{\pi/2}$ gate is implemented physically, utilizing either the robust or the constant-amplitude waveforms. The $Y_{\pi/2}$ gate is created by a simple phase advance, and the $Z_{\pi/2}$ gates are done virtually. All pulses are implemented on resonance. For resonant BARQ pulses, we use an additional phase correction (see Appendix \ref{scqc_barq_review} and Ref. \cite{CLARKEngineeringQuantumScientificComputing2021}).

The results are summarized in Fig. \ref{gst_results}, and the data are tabulated in Appendix~\ref{additional_data}. All error bars correspond to 95\% confidence intervals. The GST results were obtained using the pyGSTi Python package \cite{NIELSENProbingQuantumProcessorPerformance2020}. For direct comparison, Table \ref{gst_data_ratios} summarizes the BARQ/Constant ratios for both error metrics, and for both beam configurations. Figure~\ref{gst_results} shows that the adoption of robust pulses can boost the quantum processor performance in most cases. Noticeable improvements are evident in both error metrics, both from the figure and from the data in Table~\ref{gst_data_ratios}. 

Robust counterpropagating pulses show significant error reduction, often more than 50\% (e.g., the qubit-averaged diamond distance). The copropagating beam results, though, require a more careful look. For this geometry, only q[1] shows an improvement in diamond distance,  while all qubits except q[3] show an improvement in fidelity as a consequence of using robust pulses. The fact that q[3] does not show an improvement is likely attributable to our use of the same $s_\Omega$ for \textit{all} qubits. A more qubit-specific amplitude scale calibration routine will likely enable further error reductions. We leave this investigation for future work. Surprisingly, it appears that the adoption of robust \textit{counterpropagating} pulses is preferred over both constant and robust copropagating pulses. Further evidence for this can be seen from the data in Table~\ref{dd_infid_ratios}. 

GST attempts to make a model fit assuming Markovian errors. To quantify non-Markovianity, we count the number of standard deviations where the GST data fail to conform to a Markovian model ($N_\sigma$). Figure~\ref{gst_results}c shows an approximately 2$\times$ reduction in the value of $N_\sigma$, which indicates that implementing robust pulses leads to an effective suppression of non-Markovian errors. See Ref.~\cite{NIELSENGateSetTomography2021} for more details.

\subsection{Performance throughout quantum computer runtime}

The GST results provide important estimates of the gate error rate under the assumption of Markovian errors. During quantum computer runtime, time-dependent errors can degrade gate performance and may not be captured by the single-valued estimate provided by GST. Already, the gate set implemented with robust pulses showed lower $N_\sigma$ values on the GST model fit (see Fig.~\ref{gst_results}c), indicating the suppression of non-Markovian errors. In this section, we further investigate this observation by analyzing the results of two additional experiments.

The first experiment explores pulse robustness against motional errors, while the second assesses pulse performance in terms of repeated gate applications. The experiments are performed for three different cases: (i) drive only q[1], (ii) drive q[1] and q[2], and (iii) drive all qubits simultaneously. In all cases, both copropagating and counterpropagating beam geometries are implemented and compared. In cases (ii) and (iii), the parallel gates will be sensitive to crosstalk due to beam spillover from the neighboring ions. For the copropagating geometry, we utilize different detunings so that these effects are suppressed to first order \cite{CHOWFirstorderCrosstalkMitigationParallel2024}. The counterpropagating geometry is implemented with a global beam; therefore, such a method is not available. However, if the neighboring ions are driven by the same pulse, crosstalk will manifest as amplitude noise that can be suppressed by robust pulses.

\subsubsection{Suppressing momentum kicks with robust pulses}
\label{momentum_kicks_section}

The noise model introduced in Sec. \ref{experimental_considerations} describes errors occurring within the qubit subspace. In essence,  it assumes that the presence of the motional modes does not influence the average Rabi rate during the carrier transitions. There is clear evidence that the ion-motion coupling can significantly alter the effective Rabi strength \cite{WINELANDExperimentalIssuesCoherentQuantumstate1998, LEIBFRIEDQuantumDynamicsSingleTrapped2003}. The most characteristic example is Rabi oscillation dephasing \cite{SEMENINDeterminationHeatingRateTemperature2022}. Motion-induced decoherence is more prominent in the counterpropagating beam configuration, where momentum kicks are generated from the approximately $2\vec{k}$ $k$-vector. 

Over the course of running a trapped-ion quantum computer, the average number of phonons in each motional mode will increase. This sets an upper limit on the execution time after which the ions need to be recooled. To obtain a better grasp on how ion motion introduces errors, we approximate Eq.~\eqref{ion_chain_hamiltonian_rot} to leading order in $\vec \eta$:
\begin{align}
       \chi_{\vec n}(i \vec \eta) \approx 1 - \frac{1}{2} ||\vec \eta||_2^2 - \sum_{m=1}^{3L} n_m \eta_m^2 = 1 + \varepsilon_{\vec n}.
\end{align}
We can describe our system beyond the LD limit with
\begin{align}
    H_{\text{sys}}^{(2)} &= \sum_{\vec n} H_0(1+\varepsilon_{\vec n}) \otimes   \projector{\vec n},
    \label{beyond_ld_sys_hamiltonian}
\end{align}
where the superscript denotes second order with respect to $\eta_m$. The form of this Hamiltonian reveals that, in each phonon subspace, the qubit undergoes a rotation driven by a Rabi strength that is shifted by an amount that depends on the phonon number, which leads to Rabi oscillation dephasing~\cite{SEMENINDeterminationHeatingRateTemperature2022}. The net qubit evolution is determined by a weighted average that depends on the initial state of the motional modes. Our robust pulse is designed to mitigate such amplitude errors; therefore, the lower phonon-number subspaces will appear approximately noiseless. Although higher phonon-number sectors might translate to larger noise strengths that cannot be suppressed by our method (see Appendix \ref{scqc_barq_review}), their contribution to the averaged qubit dynamics will be small if the initial motional state assigns small probability to these sectors (for instance, if the ion chain is initially in a thermal state). We note that while our pulses modulate both amplitude and phase, constant-envelope segments with varying phase, so-called TOD sequences~\cite{VANDAMMEMotioninsensitiveTimeoptimalControlOptical2025}, have been proposed to suppress such an effect.

\begin{figure*}[htbp]
    \centering
    \includegraphics{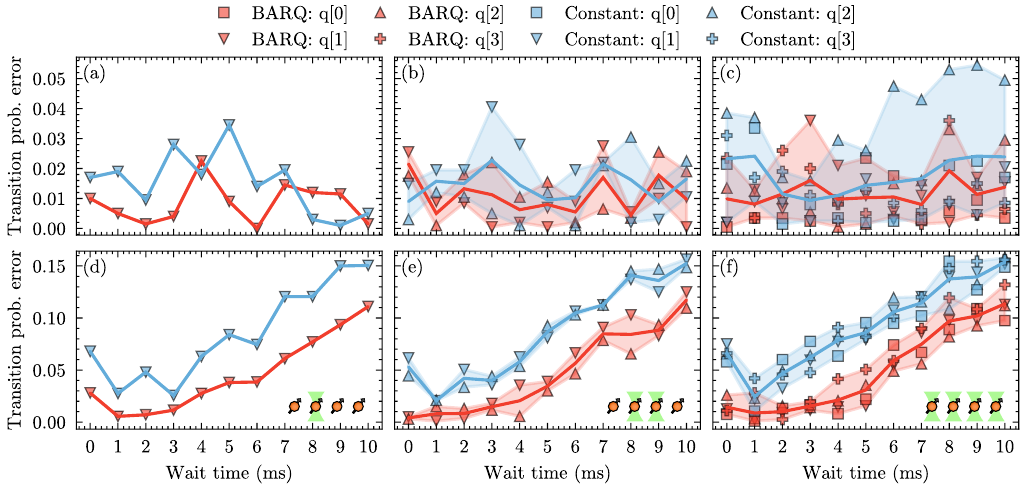}
    \caption{\textbf{Effect of motion on qubit rotation.} 
    The experimentally measured error in the probability to transition from $\ket{0}$ to $\ket{1}$ (transition prob. error $=|p-p_g|$, where $p=|\bra{1}U\ket{0}|^2$ is the experimentally measured transition probability and $p_g=|\bra{1}U_g\ket{0}|^2$ is the ideal probability) versus the wait time between preparing $\ket{0}$ and applying $U$. Here, we take $U$ and $U_g$ to be actual and ideal $X_{\pi/2}$ gates, respectively. The top row shows the results for the copropagating beam geometry, while the bottom row shows the data taken in the counterpropagating configuration. Each column shows results for one of the three cases: driving only q[1] (a,d), both q[1] and q[2] (b,e),
    and all ions (c,f).
    Panels (a)--(c) show no significant variations with respect to the wait time since the motional coupling is weak in the copropagating geometry. Panels (d)--(e) demonstrate the effect of ion motion on the qubit rotation. The constant pulses are not robust to motion-induced amplitude errors and show growing error with increasing wait times. The robust pulses maintain a smaller error up to about 3 ms, after which the error increases as the phonon number exceeds the weak-noise regime. That said, the error still remains lower than in the constant-pulse case.} 
    \label{motion_experiment}
\end{figure*}

During the execution of a quantum circuit, the ion chain will heat, and the increase in phonon number will introduce the errors discussed previously. Depending on when a quantum gate is applied (e.g., beginning of the circuit versus deep in the circuit), the gate error at different time instants will not be the same. In order to emulate this scenario, we add some wait time before the application of the gates. This will allow more phonons to be generated, and we can investigate their influence more clearly. Specifically, after all ions are sideband-cooled, we add a variable wait time ranging from zero to ten milliseconds (10 ms). We then apply an $X_{\pi/2}$ pulse and measure the excited state probability. We repeat these steps for all three aforementioned cases and for both beam geometries.

The results are summarized in Fig. \ref{motion_experiment}, which shows the absolute difference between the experimental and nominal excited state probability values. The top row depicts the data taken in the copropagating beam geometry, while the bottom row shows the results from the counterpropagating configuration. The former configuration induces weak ion-motion coupling, and naturally, the pulse behavior remains relatively steady regardless of the wait time. In contrast, the counterpropagating beams generate carrier transitions with non-zero LD parameters. As we see from the bottom row of Fig.~\ref{motion_experiment}, the constant pulse exhibits an error that grows with increasing wait times.

The robust pulses, as expected, appear motion-resistant and retain good performance for up to 3 ms. For longer wait times, their error increases but still remains lower than that of the constant pulse. Interestingly, the robust pulse causes an over-rotation (see Appendix~\ref{additional_data}, Fig.~\ref{momentum_kicks_app}). We attribute this behavior to the higher phonon-number subspaces that contribute with large noise strengths and cannot be suppressed effectively (see Appendix~\ref{scqc_barq_review}).

\subsubsection{Repetition experiment}
As in the GST experiment, various gate errors can be amplified and quantified through the repeated application of the same operation. The repetition experiment is executed for the three cases studied in the previous experiment: drive (i) only q[1], (ii) q[1] and q[2], (iii) all qubits. It offers a different perspective on the influence of non-Markovian errors. We apply the same $X_{\pi/2}$ rotation $k$ times, where $k =4k'+2, k'\in \mathbb{N} \cup \{0\}, k \le 302$, so that the qubit would always be measured in the excited state in the absence of errors. Any measured deviations from the excited state are attributed to the effects of noise. The constant pulses should exhibit the characteristic Rabi oscillation decay. Many different noise sources can produce Rabi oscillation decay when probability values are averaged. Dephasing and amplitude noise, especially due to ion-motion coupling, are two dominant error sources that our robust gates can suppress and, hence, slow the associated decay. 

\begin{figure*}
    \centering
    \includegraphics{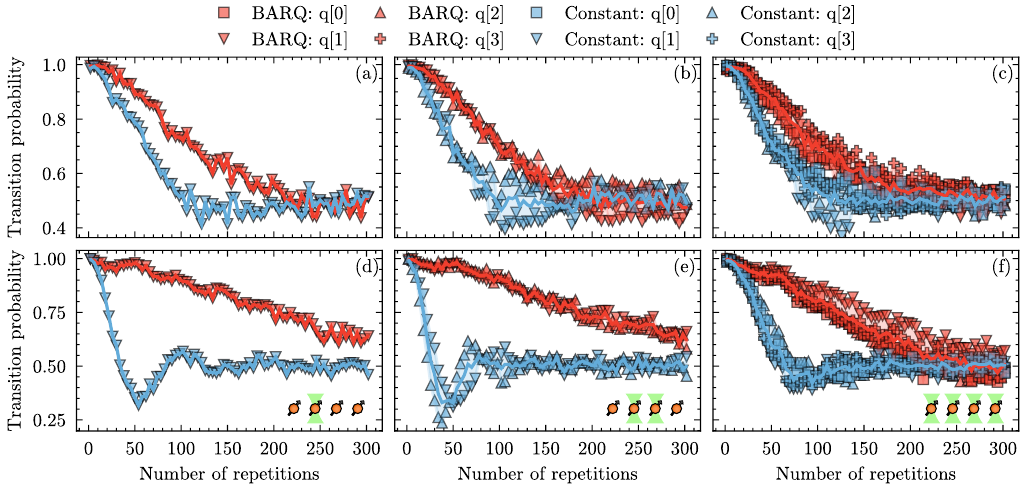}
    \caption{\textbf{Error accumulation through repeated gate application.}
   Each panel shows the experimentally measured transition probability $|\bra{1}U^k\ket{0}|^2$ versus $k$, where $k =4k'+2, k'\in \mathbb{N} \cup \{0\}, k \le 302$, and where $U$ is an $X_{\pi/2}$ gate in the absence of errors. Without noise, the probabilities should remain unity for all values of $k$ considered.
   The top row shows the results for the copropagating beam geometry, while the bottom row shows the data taken for the counterpropagating configuration. Each column corresponds to one of the three cases: driving only q[1] (a,d), both q[1] and q[2] (b,e),  
    and all ions (c,f).
    It is evident in panels (a)--(c) that the BARQ pulses produce slower decay as a function of repetition number $k$. The actual decay time is even longer since the BARQ pulse takes about 4$\times$ as long as the constant pulse. (d)--(e) The same behavior is evident for the counterpropagating configuration, where the constant pulse decays more swiftly, primarily due to the ion-motion coupling. The BARQ pulses are robust to this type of error, hence the order-of-magnitude difference in the decay rates.}
    \label{repetition_experiment}
\end{figure*}

The experimental results are depicted in Fig.~\ref{repetition_experiment}, which shows the transition probability $|\bra{1}U^k\ket{0}|^2$ as a function of $k$, where $U$ is a $X_{\pi/2}$ gate in the absence of noise. In the case of BARQ pulses, the probability decays more slowly than for the constant pulses owing to their robustness properties. While counterpropagating gates may appear to produce the slowest decay, we need to keep in mind the difference in gate times (Table~\ref{gate_time_summary}). To quantify the effect of averaging over multiple noise sources (see Appendix~\ref{prob_decay_calculations}), we fit the decaying probabilities to the function
\begin{align}
    P(t=kT_g) = \frac{1}{2} +\frac{1}{2}e^{-\frac{1}{2}(t/\tau_d)^2}\cos(\varepsilon k \pi),
\end{align}
where $\tau_d$ is a decay time constant, and $\varepsilon$ is a systematic pulse error. The decay times $\tau_d$ are summarized in Table~\ref{decay_times} and their extended version, containing the remaining data, in Table~\ref{decay_times_extended}. 

Based on the curve fit data, the copropagating BARQ pulse shows a more than 4$\times$ longer decay time compared to the constant pulse. This result indicates the successful suppression of some non-Markovian error sources. In the counterpropagating configuration, the noise-suppression benefits are more pronounced: BARQ increases the decay time by more than an order of magnitude. Despite the consistent estimates from the first two cases, a faster decay is expected when all four ions are driven. The counterpropagating pulses will be susceptible to crosstalk, and we attribute the faster decay to this effect. That said, the BARQ pulses maintain the above-noted performance difference. It is important to remind the reader that the value of $s_\Omega$ was chosen based only on q[1]. Choosing its values for each qubit while factoring in crosstalk may improve these results further.

\begin{table}
\begin{ruledtabular}
    \begin{tabular}{llccc}
 &  & \multicolumn{3}{c}{Decay time $\tau_d$ (ms)} \\
 &  & q[1] & q[1] \& q[2] & All \\
Beams & Pulse &  &  &  \\
\multirow[l]{2}{*}{Coprop.} & BARQ & 4.12 & 3.72 & 3.79 \\
 & Constant & 0.91 & 0.47 & 0.50 \\
\multirow[l]{2}{*}{Counterprop.} & BARQ & 3.35 & 3.34 & 2.08 \\
 & Constant & 0.19 & 0.16 & 0.24 \\
\end{tabular}

    \caption{\textbf{Decay times from repetition experiment.}}
    \label{decay_times}
\end{ruledtabular}
\end{table}

Before we close this section, we would like to indicate an interesting time-scale coincidence in the counterpropagating case. As we highlighted, the predicted decay times are a measure of the noise strength experienced by the gate, and we attribute the shorter decay times of the counterpropagating pulses to the additional motional noise, on top of the noise sources experienced in the copropagating case. These decay times were found to be close to 3.3 ms. Meanwhile, in the experiment discussed in Sec.~\ref{momentum_kicks_section} (that studies the effects of motional noise), we found that the BARQ counterpropagating pulses start to fail after about 3 ms of wait time.  This could indicate a connection between the two time scales. Additionally, it signifies that counterpropagating BARQ pulses might suppress all other noise sources, and their only limitation is strong motional noise, which cannot be mitigated at the control level, as discussed in Sec. \ref{momentum_kicks_section}. We leave this investigation for future work.

\section{Discussion and conclusions}
\label{conclusion}

Robust pulses are able to simultaneously suppress multiple noise sources and preserve gate quality throughout the execution of a circuit on a quantum computer. By recasting the effect of motional noise to an equivalent coherent amplitude error, we constructed dephasing- and amplitude-robust pulses to not only mitigate errors from ion motion but also errors due to laser-beam amplitude and frequency fluctuations. We rigorously characterized a four-qubit trapped-ion register using gate set tomography and carried out experiments that amplify noise in order to assess the gate performance in a realistic setting. Despite using a minimal calibration routine—tuning based only on a single qubit (q[1]) at the start of a five-month experimental period—our robust pulses provided significant performance benefits across the majority of the register.

We centered our analysis on the diamond distance as a gate-error metric given its important role in fault-tolerance thresholds. Our experiments reach a rather counterintuitive conclusion. It appears that robust counterpropagating pulses (which couple with motion strongly) can outperform copropagating pulses, including ones that dynamically suppress noise. Under this metric, we reach counterpropagating gate errors as low as $3.59 \cdot 10^{-3}$, a value that establishes a diamond distance reference for this beam geometry and is merely one order of magnitude larger than the best reported single-qubit \textit{microwave} pulse error-rate \cite{BLUME-KOHOUTDemonstrationQubitOperationsRigorous2017}. Although robust counterpropagating pulses are expected to outperform constant copropagating pulses because they suppress multiple noise sources, the fact that even robust copropagating pulses sometimes show higher error rates warrants further investigation. In particular, this will require a better understanding of the relevant noise sources and the role of gate time (for reference, the counterpropagating pulses are approximately 2.5 times faster than the copropagating ones).

The conclusion that robust counterpropagating pulses can outperform constant copropagating pulses is supported also when using the average gate infidelity as a metric. Using the gate set tomography model fit results, we estimate counterpropagating gate infidelities as low as $1.36 \cdot 10^{-3}$ (99.86\% average gate fidelity) which is roughly half of the constant copropagating pulse result. For comparison to the existing literature, we performed randomized benchmarking experiments based on the procedure in Ref. \cite{WRIGHTBenchmarking11qubitQuantumComputer2019}. We found that the maximum counterpropagating pulse fidelity estimated from RB is 99.87\% (see Table \ref{rb_data}) while Ref. \cite{WRIGHTBenchmarking11qubitQuantumComputer2019} reported fidelities as high as 99.64\%. In general, all four qubits showed greater RB fidelity values than the ones reported in Ref. \cite{WRIGHTBenchmarking11qubitQuantumComputer2019}. The differences between these values likely arise from two main factors. First, the utilized gates are different: our setup uses $X_{\pi/2}$, $Y_{\pi/2}$, and $Z_{\pi/2}$, whereas Ref. \cite{WRIGHTBenchmarking11qubitQuantumComputer2019} uses only $X_{\pi/2}$ and $Y_{\pi/2}$. This leads to different Clifford gate compilations, and in Ref. \cite{WRIGHTBenchmarking11qubitQuantumComputer2019} additional $\pi$-pulses along the $X$, $Y$, and $Z$ axes are explicitly included in the RB sequence. In our case, these gates may instead appear as composites of robust native gates, potentially reducing the error of the composed $\pi$-pulse. Second, the experimental setup in Ref. \cite{WRIGHTBenchmarking11qubitQuantumComputer2019} contains thirteen ions (eleven are utilized as qubits) versus four in our experiments. For the larger number of qubits, the existence of additional motional modes may increase the estimated gate error. In this context, RB shows reduced sensitivity against coherent errors and might not be appropriate to capture motional noise effects which are prominent in the counterpropagating geometry. Overall, RB measures the average error across random gate sequences \cite{CARIGNAN-DUGASRandomizedBenchmarkingExperimentsGateset2018}, but it does not reveal how well individual gates perform. Moreover, even if the RB infidelity is very small and treated as the average native-gate error, it may still be insufficient for fault tolerance, because the diamond-distance error can scale as the square root of the infidelity \cite{SANDERSBoundingQuantumGateError2015, WALLMANRandomizedBenchmarkingConfidence2014, WALLMANNoiseTailoringScalableQuantum2016}.

These findings realize the goal of reducing experimental complexity by paving the way to eliminating the need for copropagating beam geometries entirely. While nearly all qubits showed improvements, the performance of qubits at the edges of the device suggests that further error reduction is possible if we calibrate for each qubit individually. For systems where copropagating geometries still show advantage, we expect that robust pulses can help mitigate fluctuating errors and boost performance to even lower than the previously reported (RB-derived) $10^{-5}$ laser-driven gate infidelity~\cite{RANSFORDHelios98qubitTrappedionQuantum2025, GAEBLERHighFidelityUniversalGateSet2016}. The adoption of robust pulses can help both reduce and sustain low single-qubit error rates while improving the performance gains of fault-tolerant quantum computing experiments \cite{RYAN-ANDERSONRealizationRealTimeFaultTolerantQuantum2021, NGUYENDemonstrationShorEncodingTrappedIon2021, MOSESRaceTrackTrappedIonQuantumProcessor2023}.

\section*{Acknowledgments}
The authors would like to thank Corey Ostrove for the extremely helpful discussions about gate set tomography and pyGSTi. E.P. thanks Filippos Dakis for extensive discussions on data representation. 

SEE acknowledges support from the Department of Energy, Grant No. DE-SC0024488, which supported the experimental design (in particular incorporating the effect of spectator qubits), the experimental testing of pulses, and data analysis including tomography and randomized benchmarking simulations and GST plots. EB acknowledges support from the Office of Naval Research,
grant no. N00014-25-1-2125, which enabled the development and experimental evaluation of robust pulse methods for ion control, including noise characterization, tomography and data-analysis tools, and coordination of experiments probing the potential advantages of robust control.
This research was supported by the U.S. Department of Energy, Office of Science, Office of Advanced Scientific Computing Research Quantum Testbed Program. Sandia National Laboratories is a multi-mission laboratory managed and operated by National Technology \& Engineering Solutions of Sandia, LLC (NTESS), a wholly owned subsidiary of Honeywell International Inc., for the U.S. Department of Energy’s National Nuclear Security Administration (DOE/NNSA) under contract DE-NA0003525.  SAND2026-21969O.  This written work is coauthored by an employee of NTESS. The employee, not NTESS, owns the right, title and interest in and to the written work and is responsible for its contents. Any subjective views or opinions that might be expressed in the written work do not necessarily represent the views of the U.S. Government. The publisher acknowledges that the U.S. Government retains a non-exclusive, paid-up, irrevocable, world-wide license to publish or reproduce the published form of this written work or allow others to do so, for U.S. Government purposes. The DOE will provide public access to results of federally sponsored research in accordance with the DOE Public Access Plan.  

\section*{Data availability}
The data that support the findings of this work are openly available in the GitHub repository \url{https://github.com/evpiliouras/barq\_on\_trapped\_ions}.

\bibliography{trappedionrefs}

\appendix
\setcounter{table}{0} 
\renewcommand{\thetable}{A\arabic{table}}
\section{Derivations for the ion chain system}
\label{ion_derivations}
In this section, we provide the calculations that support our system description and noise models. The Hamiltonian describing an $L$-ion chain interacting with electromagnetic fields is written as \cite{WINELANDExperimentalIssuesCoherentQuantumstate1998}
\begin{align}
    H = \sum_{m=1}^{3L} \omega_m a_m^\dagger a_m 
    -\frac{\omega_0}{2}\sum_{l=1}^L \sigma_{z, l} -\sum_{l=1}^L\hat \epsilon_l \cdot \vec E_l(\vec r_l) .  
\end{align}
The first term contains the $3L$ ion chain modes of oscillation given by the frequencies $\omega_k$ and the respective phonon annihilation (creation) operators $a_k (a_k^\dagger)$. We note that the motional part refers to the collective ion motion as a result of the Coulombic repulsion and not the individual ion motion. This will become an important distinction as we proceed with the derivations. We assume that the motional quantization has occurred about the equilibrium point of the chain \cite{JAMESQuantumDynamicsColdTrapped1998}. The second term defines the qubit's energy splitting $\omega_0$ and $\sigma_{z,l}$ is the Pauli $\sigma_z$ operator acting on the $l$-th ion. The last term introduces the laser fields $\vec E_l$ coupling with the $l$-th ion's dipole moment as a function of its position $\vec r_l$. The results are structurally the same regardless of whether the gates are implemented through microwaves or using laser-driven Raman transitions \cite{WINELANDExperimentalIssuesCoherentQuantumstate1998, LEIBFRIEDQuantumDynamicsSingleTrapped2003}.

We will drop the index $l$ addressing each ion.
We move to the rotating frame (laser-field interaction picture) according to the free evolution, using the transformation
\begin{align}
    R(t) = \text{exp}\left(-it\left[\sum_{m=1}^{3L} \omega_m a_m^\dagger a_m 
    -\frac{\omega_0}{2}\sigma_z\right]\right).
\end{align}
The field interaction term can be written as
\begin{align}
    -\hat \epsilon \cdot \vec E(\vec r)  =\Omega \cos(\vec k \cdot \vec r-\omega_d t +\Phi) (\sigma + \sigma^\dagger), 
\end{align}
where $\omega_d$ is the driving frequency, $\Omega$ is the Rabi strength and $\Phi$ the phase of the field. The atomic lowering and raising operators are defined as $\sigma$ and $\sigma^\dagger$, respectively. We note the following identities:
\begin{align}
R^\dagger& (\sigma+ \sigma^\dagger) R = \sigma e^{-i\omega_0t}+ \sigma^\dagger e^{i\omega_0t}, \\
    R^\dagger& \left(\vec k \cdot \vec r \right)  R  = 
    R^\dagger \left(\sum_{m=1}^{3L} b_{m}  \, \vec k \cdot \vec \xi_m \right) R \nonumber\\
    &= \sum_{m=1}^{3L} b_{m} R^\dagger \left[ \vec k \cdot \hat e_m \,\xi_{m, \text{ ZPF}}  (a_m + a_m^\dagger) \right] R  \nonumber\\
 &= \sum_{m=1}^{3L} \eta_m (a_m e^{-i\omega_m t} + a_m^\dagger e^{i\omega_m t}),
\end{align}
where $\eta_m = b_m \vec k \cdot \vec \xi_{m, \text{ ZPF}}$ is the Lamb-Dicke parameter associated with mode $m$, and $b_m$ is the projection of the position vector $\vec r$ along the normal mode vector $\vec \xi_m$. Upon motion quantization, we have $\vec \xi_m= \vec \xi_{m, \text{ ZPF}}(a_m + a_m^\dagger) = \hat e_m\, \xi_{m, \text{ ZPF}}(a_m + a_m^\dagger)$ with $\xi_{m, \text{ ZPF}}$ denoting the zero-point fluctuations of the $m$-th mode and $\hat e_m$ the unit vector associated with that mode. In order to further study the effect of the field interaction term, we first notice that
\begin{align}
    &R^\dagger \exp{(i \vec k \cdot \vec r)} R = \exp{(i R^\dagger \left[\vec k \cdot \vec r \right ]R)} \nonumber\\
    &= \exp{\left[i \sum_{m=1}^{3L} \eta_m (a_m e^{-i\omega_m t} + a_m^\dagger e^{i\omega_m t}) \right ]} \nonumber \\
    &=\bigotimes_{m=1}^{3L} \exp\left[i\eta_m (a_m e^{-i\omega_m t} + a_m^\dagger e^{i\omega_m t}) \right].
    \label{rotating_cosine_term}
\end{align}
Towards simplifying the interaction picture Hamiltonian, the constituent oscillating frequencies of this particular term can be identified with the help of the occupation number states $a_m^\dagger a_m \ket{n_m} =n_m \ket{n_m}$. We insert their resolution of identity and write:
\begin{align}
    \exp\left[i\eta_m (a_m e^{-i\omega_m t} + a_m^\dagger e^{i\omega_m t}) \right] \nonumber\\
    = \sum_{n_m, n'_m} \bra{n_m} \exp\left[i\eta_m (a_m  + a_m^\dagger) \right] \ket{n'_m}  \ket{n_m}\!\!\bra{n'_m} \nonumber\\
    \times \exp(i(n_m -n'_m)\omega_m t).
\end{align}
If we drive on resonance with the electronic transition ($\omega_d=\omega_0)$ while expanding the cosine terms into exponentials, terms that exchange motional quanta with qubit excitations will appear as fast-rotating terms. With the application of the rotating wave approximation (RWA), we can write
\begin{align}
    \exp\left[i\eta_m (a_m e^{-i\omega_m t} + a_m^\dagger e^{i\omega_m t}) \right] \nonumber\\ \approx \sum_{n_m} \chi_{n_m}(i \eta_{m})  \projector{n_m},
\end{align}
with $\chi_{\rho}(\alpha) = \text{tr}(D(\alpha)\rho)$ and $D(\alpha) = e^{\alpha a^\dagger -\alpha^* a}$. The characteristic function for every mode containing $n_m$ motional quanta becomes
\begin{align}
    \chi_{n_m}(i \eta_{n_m}) = \bra{n_{m}} 
    \exp[i\eta_{m}(a_{m} + a_{m}^\dagger)]
    \ket{n_{m}} \nonumber\\
    = e^{-\frac{\eta_{m}^2}{2}}L_{n_m}(\eta_{m}^2),
\end{align}
where $L_{n_m}$ are the Laguerre polynomials. We can then apply the RWA on Eq. \eqref{rotating_cosine_term} to obtain
\begin{align}
    \bigotimes_{m=1}^{3L} \exp\left[i\eta_m (a_m e^{-i\omega_m t} + a_m^\dagger e^{i\omega_m t}) \right] \nonumber\\
    = \bigotimes_{m=1}^{3L} \sum_{n_m} \chi_{n_m}(i \eta_{n_m}) \projector{n_m}\nonumber \\
    =\sum_{\vec n} \chi_{\vec n}(i \vec \eta) 
    \projector{\vec{n}},
\end{align}
where we defined $\vec n = [n_1 \, n_2 \dots n_{3L}]^T$ as the vector of motional occupation numbers, as well as the associated quantum state
\begin{align}
    \ket{\vec{n}} = \ket{n_1\,n_2\dots n_{3L}},
\end{align}
and $\vec \eta$ as a vector that contains the LD parameters corresponding to each mode. The coefficients that depend on the LD parameters and the motional occupation numbers are understood as the product
\begin{align}
    \chi_{\vec n}(i \vec \eta) = \prod_{m=1}^{3L} \chi_{n_m}( i\eta_m).
\end{align}
The laser-field interaction picture Hamiltonian becomes
\begin{align}
    H_{I, R} = \frac{\Omega}{2}(\sigma^\dagger e^{i\Phi} + \sigma e^{-i\Phi}) \otimes \sum_{\vec n} \chi_{\vec n}(i \vec \eta) \projector{\vec n}.
\end{align}
We can now define the single-qubit control Hamiltonian as
\begin{align}
     H_0(t) = \frac{\Omega(t)}{2}[\cos\Phi(t)\sigma_x + \sin\Phi(t)\sigma_y],
\end{align}
and write the carrier transition as
\begin{align}
    H_{I, R} =  \sum_{\vec n} \chi_{\vec n}(i \vec \eta) H_0(t) \otimes \projector{\vec n}.
\end{align}
Since every mode state $\ket{\vec n}$ is orthogonal to the rest and time-independent, we can write the propagator as 
\begin{align}
    U_{I, R} = \sum_{\vec n} U_{\vec n} \otimes \projector{\vec n}
    \label{general_propagator}
\end{align}
where every phonon subspace is described by a separate Schr\"{o}dinger equation: 
\begin{align}
    i \dot U_{\vec n} = \chi_{\vec n}(i \vec \eta) H_0(t) U_{\vec n}.
\end{align}

Before we conclude this section, it is important to gain some intuition on how the Rabi-modulating function $\chi_{\vec n}(i \vec \eta)$ affects the single-qubit rotations in each subspace. We expand the term $\chi_{\vec n}(i \vec \eta)$ with respect to $\vec \eta$ to find
\begin{align}
    \chi_{\vec n}(i \vec \eta) \approx 1 - \frac{1}{2} ||\vec \eta||_2^2 - \sum_{m=1}^{3L} n_m \eta_m^2 = 1 + \varepsilon_{\vec n}.
\end{align}
When the parameter $\vec \eta \to \vec 0$, which is the case for the copropagating beam configuration, the evolution is identical within each phonon subspace. When counterpropagating beams are used, the momentum kicks will cause each phonon subspace to execute a rotation with modified Rabi strength, with a value dependent on the number of motional quanta. This effect is an off-resonant phenomenon; the qubit-field detuning is far away from the ion-chain's frequencies for excitations to be exchanged, but the presence of the motional degrees of freedom modifies the energy structure. This situation is loosely similar to the case of the dispersive coupling in the atom-cavity setup.

We readily see from Eq. \eqref{general_propagator} that the final qubit evolution is obtained from a weighted average that depends on the ion-chain initial state. Here, we assumed that pulses are performed on a smaller timescale compared to the rate at which motional quanta accumulate. A more detailed treatment would entail an open quantum system description.

\section{Overview of SCQC and BARQ}
\label{scqc_barq_review}
In this section, we will briefly discuss the robust pulse design procedure using SCQC and dive deeper into curve optimization using BARQ. 
\subsection{Suppressing noise with SCQC}
While the problem stated and developed throughout this section is for single-qubit control, SCQC is already extended for higher dimensions \cite{BUTERAKOSGeometricalFormalismDynamicallyCorrected2021}. We consider the control problem statement in Ref. \cite{PILIOURASAutomatedGeometricSpaceCurve2026} as
\begin{align}
    i\dot U_0 = H_0 U_0,\qquad U_0(T_g) = U_g,
\end{align}
where $H_0$ is the noise-free (ideal) control Hamiltonian and $U_g$ the ideal quantum operation to be achieved at some final time $T_g$. In the context of this work, $H_0$ is recovered as the carrier transition in the ion's electronic and motional degrees of freedom rotating frame. We additionally consider arbitrary error entering as $H = H_0 +H_\text{n}$ and decompose the evolution as $U=U_0U_I$ with
\begin{align}
    i\dot U_I = (U_0^\dagger H_\text{n} U_0)U_I.
\end{align}
All quantum control design methods will need to achieve both the desired gate (gate-fixing) and exhibit noise-insensitivity (noise-robustness) in the presence of $H_\text{n}$. The former is quantified with the average gate fidelity defined as:
\begin{align}
   \mathcal{F}(M,I_d) = \mathcal{F}(M) = \frac{\text{tr}(MM^\dagger)+|\text{tr}(M))|^2}{d(d+1)} ,
\end{align}
with $M = U_0(T_g)U_g^\dagger$, while the latter by the quantity
\begin{align}
  \left|\left| \frac{1}{T_g} \int_0^{T_g} dt\, U_0^\dagger (T_g H_\text{n})U_0 \right|\right|_2^2.
  \label{pulse_robustness_goal}
\end{align}
This quantity is retrieved through the first-order Magnus expansion term \cite{BLANESMagnusExpansionItsApplications2009} or the average Hamiltonian theory \cite{BRINKMANNIntroductionAverageHamiltonianTheory2016}.

SCQC starts the design procedure backwards by aiming to identify the common characteristic of all dephasing robust pulses, i.e., controls that suppress noise of the form $H_\text{n} = \frac{\delta_z}{2}\sigma_z$. Based on the relevant quantity defined in Eq. \eqref{pulse_robustness_goal}, we write the Heisenberg picture of the error operator $\sigma_z$ in terms of the Pauli operators as
\begin{align}
    \int_0^{t} dt' \, U_0^\dagger \sigma_z U_0 = \vec r (t) \cdot \sigvec,
    \label{scqc_main_eq_app}
\end{align}
where $\sigvec = \begin{bmatrix}
    \sigma_x & \sigma_y & \sigma_z
\end{bmatrix}^T$ is a vector that contains the Pauli operators and $\vec r(t)$ is a three-dimensional space curve with time-dependent components. It is clear that for robustness to be achieved to first-order with respect to $\delta_z$, we need
\begin{align}
    \vec r(T_g) = \vec r(0).
\end{align}
In other words, designing a closed curve will correspond to a set of controls that suppress frequency errors.
We will consider the general single-qubit Hamiltonian
\begin{align}
H_0(t) = \frac{\Omega(t)}{2}[\cos\Phi(t)\sigma_x + \sin\Phi(t)\sigma_y] + \frac{\Delta(t)}{2}\sigma_z,
\end{align}
with the same fields defined in the main text, but also allow for arbitrary detuning $\Delta$.
In order to recover the controls, we differentiate Eq. \eqref{scqc_main_eq_app} with respect to $t$ and find that $\vec T \equiv \dot {\vec r}$, where $\vec T$ is the tangent vector of the curve $\vec r$. In differential geometry terms, this equation tells us that time coincides with the length along the curve (arclength parameterization). Subsequent differentiations and algebraic manipulations (see Ref. \cite{PILIOURASAutomatedGeometricSpaceCurve2026} for more details) allow us to arrive at the set of equations
\begin{align}
    \dot{\vec T} \cdot \sigvec = -\Omega\,(\sin\Phi\, \hat x - \cos\Phi\, \hat y)\cdot \sigvec_{U_0}  \label{scqc:dot_t_eq},\\
   (\vec T \times  \dot{\vec T}) \cdot \sigvec = -\Omega\,(\cos\Phi\, \hat x+ \sin\Phi\, \hat y  )\cdot \sigvec_{U_0}  \label{scqc:doubledot_t_eq},
\end{align}
where $\sigvec_{U_0} \equiv U_0^\dagger \,\sigvec\, U_0$. 

One can define an orthonormal moving frame attached at each point along the curve, called the Frenet-Serret (FS) frame, that can help us both recover the control fields and determine the final quantum operation, while avoiding having to solve the Schr\"{o}dinger equation. The FS frame is created by the tangent vector and the mutually orthogonal vectors, normal vector $\vec N = \dot{\vec T}/||\dot{\vec T}||_2$ and binormal vector $\vec B  = \vec T \times \vec N$. The normal vector's relative rates of change define two geometric quantities that characterize the curve up to rigid rotations and translations, namely the curvature $\kappa(t)$ and the torsion $\tau(t)$. They are defined as
\begin{align}
    \kappa(t) = -\dot {\vec  N} \cdot \vec T ,\\
    \tau(t) =  \dot {\vec  N} \cdot \vec B.
\end{align}
Further time-differentiation of Eq. \eqref{scqc:doubledot_t_eq} and operator norm calculations lead to the control fields as
\begin{align}
    \Omega(t) &= \kappa(t) \label{envelope_cond}, \\
    \dot \Phi(t) - \Delta(t) &=\tau(t).
    \label{phi_delta_cond}
\end{align}
Some special cases where the curvature vanishes (inflection points) are handled in detail in Ref. \cite{PILIOURASAutomatedGeometricSpaceCurve2026}.

For a broader understanding of the SCQC evolution, it is useful to use the adjoint representation 
\begin{align}
    R_U^{ij} = \frac{1}{2}\text{tr}(U^\dagger\sigma_iU\sigma_j).
\end{align}
The adjoint representation can be recognized as the ~~$(d^2-1) \times (d^2-1)$ sub-block of the Pauli Transfer Matrix (PTM) \cite{PILIOURASAutomatedGeometricSpaceCurve2026, NIELSENGateSetTomography2021}. When expressed in SCQC terms, it is written as
\begin{align}
    R_{U_0}(t) = R_Z(\Phi(t))R_F(t) R_F^T(0),
    \label{adjoint_rep_scqc}
    \end{align}
where:
\begin{align}
    R_{F}(t) = \begin{bmatrix}
        -\vec B\hphantom{-} \\
        \hphantom{-}\vec N\hphantom{-} \\
        \hphantom{-}\vec T \hphantom{-}
    \end{bmatrix},
    \label{rotmatF_def}
\end{align}
\begin{align}
    R_Z(\Phi(t)) =  \begin{bmatrix}
        \cos\Phi(t) & -\sin\Phi(t) & 0 \\
        \sin\Phi(t) & \hphantom{-}\cos\Phi(t) &0 \\
        0&0&1
    \end{bmatrix}.
\end{align}
The average gate fidelity can be written as
\begin{align}
   \mathcal{F}_g(U_0(T_g)) = \frac{d+1+\text{tr}(R_g^TR_{U_0}(T_g))}{d(d+1)}
\end{align}
which becomes unit when $R_{U_0}(T_g)=R_g$. Essentially, fixing the relative orientation of the FS frame between $t=0$ and $t=T_g$, we can achieve the desired gate up to a Z-rotation. The remaining Z-rotation is removed either with a constant detuning or using a virtual-Z gate. If no such options are available, this must be handled in the curve design step.

Before we conclude the SCQC review, it is important to mention that our noise-suppression strategy is valid as long as the noise is weak and the Magnus expansion converges. The latter is guaranteed to happen as long as
\begin{align}
    \int_0^{T_g} dt ||H_{I}||_2 < \pi.
\end{align}
This is an important point for our discussion in the main text in Sec. \ref{momentum_kicks_section}, where we investigate the suppression of amplitude errors induced by the ion motion.
From Eq.~\eqref{beyond_ld_sys_hamiltonian}, we understand that each phonon subspace enjoys amplitude-error suppression provided that
\begin{align}
    \left(\int_{0}^{T_g} dt\,  \Omega (t) \right )  |\varepsilon_{\vec n}| < 2\pi .
\end{align}
Depending on the phonon subspace occupation number, amplitude errors in phonon number subspaces with numbers defined outside the region of the cross polytopes given by 
\begin{align}
    \left | \frac{1}{2} ||\vec \eta||_2^2 + \sum_{m=1}^{3L} n_m \eta_m^2\right| <2\pi \left[ \int_{0}^{T_g} dt\,  \Omega (t)   \right]^{-1}
\end{align}
are not guaranteed to be suppressed by our theory.
That said, depending on the values of the function $p_{\vec n}=\bra{\vec n} \rho_{\text{motion}} \ket{\vec n}$ where $\rho_{\text{motion}}$ indicates the initial ion chain state, contributions from some phonon subspaces might not even appear so our strategy can remain valid for cases where the probability becomes smaller for higher energy states (e.g. thermal states).

\subsection{Optimizing curves using BARQ}
As discussed in the previous section, the adjoint representation provides the geometric guidance on how to construct gates up to a Z-rotation. BARQ uses this information to provide an automated method for constructing gates that a priori have unit fidelity in the noise-free case and require optimization only for the robustness properties. This solves the usually unavoidable tradeoff between fixing the target gate and achieving noise insensitivity \cite{PILIOURASAutomatedGeometricSpaceCurve2026}. In SCQC, time is the arclength parametrization since $||\dot {\vec r}(t)||_2=1 \,\,\forall t$. This would appear as a restrictive feature when choosing a curve parametrization, but in fact, one can use a simple change of variables to address any type of curve. If $x$ is the non-arclength parameter, we can express the time variable as
\begin{align}
    dt/dx = \int_0^x dx' ||dr(x')/dx'||_2.
\end{align}
BARQ uses the Bézier curve parameterization, which is defined through the equation
\begin{align}
    \vec r(x) = \sum_{j=0}^{n} \vec w_j g_{j,n}(x),\qquad x \in [0,1], 
\end{align}
where the curve is formed using $n+1$ control points $\vec w_j$. The basis functions are the Bernstein polynomials given by
\begin{align}
    g_{j,n}(x) = \begin{pmatrix}
        n \\ j
    \end{pmatrix} x^j(1- x)^{n-j}.
\end{align}
Utilizing the analytic equation of the adjoint representation, one can encode the gate inside the control point arrangement and optimize their position only for robustness and pulse properties. Additionally, one can create a closed curve simply by defining equal endpoints, i.e., $\vec w_0=\vec w_{n}$. Having made the above steps, we are able to construct an informed ansatz that guarantees some properties upfront.

The optimization goals are implicitly defined through objective functions that enforce the required conditions as they reduce to zero. This is the general approach when multiple goals are targeted or analytical equations are not available to impose them exactly. Similar to the case of the examples in Ref. \cite{PILIOURASAutomatedGeometricSpaceCurve2026}, we use the total objective function
\begin{align}
    J_{\text{BARQ}} = \Big\|\int_0^{T_g} dt \,\vec T \times \dot{\vec T}\;\Big\|_2^2 + 10^{-2} T_g \max_t\{\kappa(t)\} .
\end{align}
The first term enforces the zero-area tangent condition that guarantees amplitude-error robustness, while the second term minimizes the dimensionless Rabi rate maximum, a quantity that is crucial for the experimental gate duration. The relative weight $10^{-2}$ denotes a preferential optimization towards amplitude robustness. The curve was optimized with a total of 10 free points. More details can be found in the ``Results" section of \cite{PILIOURASAutomatedGeometricSpaceCurve2026}. The curve was optimized using the \texttt{qurveros} Python package \cite{PILIOURASQurverosSCQCBARQImplementation2025}.

The final element in implementing the desired quantum operation is the handling of the final Z-rotation. The angle of rotation depends on $\Phi(T_g)$ and any additional rotation introduced as a result of the orientation between the initial- and final-time FS frames. This part can be handled either with a constant detuning value that allows the total angle to be $2k\pi, k \in \mathbb{Z}$, or a simple virtual Z-rotation can be implemented after the waveform is applied. The latter was adopted in our work. Alternatively, one can add an additional constraint to the total objective function.

\section{Brief sketch of probability decay}
\label{prob_decay_calculations}

In this section, we motivate the existence of probability decay when we apply our gate multiple times. We assume that the gate has been applied $k$ times and write the evolution operator recursively as
\begin{align}
    U(k T_g) = U_0^{[k]}U_I^{[k]} U([k-1]T_g).
\end{align}
We assume that each pulse application lasts for time $T_g$ and yields a quantum gate of the form $U_0^{[k]}U_I^{[k]}$. The entire evolution is captured by $U(kT_g)$. We follow the same reasoning as Refs. \cite{KHODJASTEHDynamicallyErrorCorrectedGatesUniversal2009, MAVADIAExperimentalQuantumVerificationPresence2018}. We need to isolate the effect of the total evolution from its noisy parts, and we proceed by noticing that
\begin{align}
    &U(k T_g) = U_0^{[k]}U_I^{[k]} U([k-1]T_g) \nonumber\\
    &=U_0^{[k]}U_I^{[k]} {U_0^{[k]}}^\dagger U_0^{[k]}U([k-1]T_g)\nonumber\\
    &= \tilde{U}_I^{[k]} U_0^{[k]} U([k-1]T_g) = \nonumber\\ 
    &= \tilde{U}_I^{[k]} U_0^{[k]}U_0^{[k-1]} U_I^{[k-1]}U([k-2]T_g) \nonumber \\
    &=\tilde{U}_I^{[k]} \mathcal{C}_0^{[k:1]} U_I^{[k-1]}U([k-2]T_g)
\end{align}
where we defined
\begin{align}
    &\mathcal{C}_0^{[k:k']} =  U_0^{[k]}U_0^{[k-1]}\cdots U_0^{[k-k']}, k'\le k, k>0,\\
    &\tilde{U}_I^{[k-k']} = \mathcal{C}_0^{[k:k']} {U}_I^{[k-k']}{\mathcal{C}_0^{[k:k']}}^\dagger.
\end{align}
We now repeat the same procedure using $\mathcal{C}_0^{[k:k']}$ and its conjugate to transform every ${U}_I^{[k-k']}$ and obtain
\begin{align}
    U(kT_g) = \left[\tilde{U}_I^{[k]}\tilde{U}_I^{[k-1]}\cdots \tilde{U}_I^{[0]}\right] \mathcal{C}_0^{[k:k]}.
    \label{final_evol_noisy}
\end{align}
Considering $U_I^{[k-k']} = \exp(-i\theta_{k-k'} \,\vec n_{k-k'} \cdot \sigvec)$, the action of $\mathcal{C}_0^{[k:k']}$ is to rotate the vector $\vec n_{k-k'}$ according to the accumulated rotation of $U_0^{[k]}$. If we approximate Eq. \eqref{final_evol_noisy} to first order of $\theta_{k-k'}$, we can
identify a similar random-walk interpretation \cite{MAVADIAExperimentalQuantumVerificationPresence2018}. By virtue of the central limit theorem, we postulated a Gaussian decay characterized by $\tau_d$ in the main text. Note that $\mathcal{C}_0^{[k:k]}$ accumulates all ideal rotations up to $k$. Since the qubit is measured at repetition numbers where $k =4k'+2, k'\in \mathbb{N} \cup \{0\}$, $\mathcal{C}_0^{[k:k]}$ = $X_\pi$ at all times. Before averaging, the excited state probability will be
\begin{align}
    P(t=kT_g) = \left| \Bra{1}  \tilde{U}_I^{[k]}\tilde{U}_I^{[k-1]}\cdots \tilde{U}_I^{[0]} \Ket{1}\right|^2
\end{align}
and averaging over the stochastic noise variables will introduce the proposed decay.

\section{Additional data and data analysis}
\label{additional_data}
This section includes additional data that complements the discussion of the main text and briefly describes the data aggregation in the GST experiment. We performed several GST experiments on qubits q[1] \& q[2] and a set of all beam geometry/pulse combinations on the edge ions, qubits q[0] \& q[3]. The qubit ordering is different than the one commonly used in the QSCOUT system \cite{YALERealizationCalibrationContinuouslyParameterized2025}. The amount of time required to obtain these data can be prohibitive for extensive characterization of all qubits. For example, running the 733 circuits for single qubit GST, each with varying numbers of gates, 200 times each took roughly 30 minutes for the fastest gates (constant pulses in counterpropagating configuration) and up to 70 minutes for the slowest gates (BARQ pulses in copropagating configuration). From the collected data, some entries are removed from the pool due to prohibitively large diamond distance error bars. 

The data and their analysis are openly available in the GitHub repository \url{https://github.com/evpiliouras/barq\_on\_trapped\_ions}. The data are aggregated by finding the mean values over all experiments for each combination of beam geometry, qubit, and pulse, using the Pandas library \cite{THEPANDASDEVELOPMENTTEAMPandasdevPandasPandas2024}. The data are summarized in Table \ref{gst_results_table} and additional metric ratios are reported in Table \ref{dd_infid_ratios}.

\begin{figure*}
    \centering
    \includegraphics{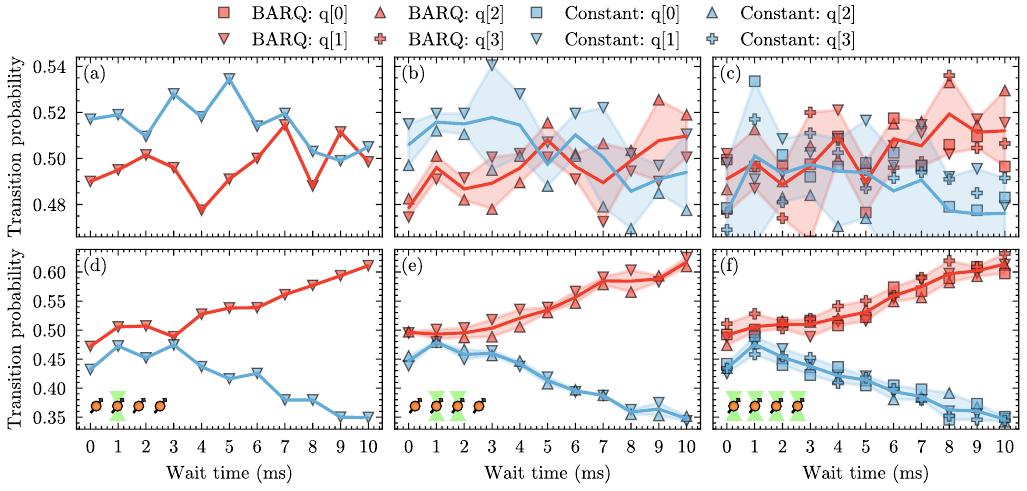}
    \caption{An alternative view for Fig. \ref{motion_experiment}. Here, we plot the measured probability values to show that robust pulses tend to over-rotate the qubit with increasing wait times. The constant pulse follows the expected under-rotation as discussed in the main text.}
    \label{momentum_kicks_app}
\end{figure*}

In Table \ref{decay_times_extended}, we provide an extended table of the main text Table \ref{decay_times}. We include the predicted pulse error and the fit error (sum squared differences) to ensure that our fit is as close to the experimental data as possible. Finally, Fig. \ref{momentum_kicks_app} plots the probability value (instead of the absolute difference with respect to the nominal value 0.5) to show that the robust pulses cause over-rotations for increasing values of wait times. 

\begin{table*}
\begin{ruledtabular}
        \begin{tabular}{llccccccccc}
 &  & \multicolumn{3}{c}{Decay time $\tau_d$ (ms)} & \multicolumn{3}{c}{Pulse error $\varepsilon$} & \multicolumn{3}{c}{Fit error} \\
 &  & q[1] & q[1] \& q[2] & All & q[1] & q[1] \& q[2] & All & q[1] & q[1] \& q[2] & All \\
Beams & Pulse &  &  &  &  &  &  &  &  &  \\
\multirow[l]{2}{*}{Coprop.} & BARQ & 4.12 & 3.72 & 3.79 & $5.42 \cdot 10^{-6}$ & $5.91 \cdot 10^{-6}$ & $5.71 \cdot 10^{-7}$ & $6.74 \cdot 10^{-4}$ & $2.44 \cdot 10^{-4}$ & $5.67 \cdot 10^{-4}$ \\
 & Constant & 0.91 & 0.47 & 0.50 & $4.85 \cdot 10^{-3}$ & $1.08 \cdot 10^{-5}$ & $6.06 \cdot 10^{-6}$ & $7.23 \cdot 10^{-4}$ & $4.47 \cdot 10^{-4}$ & $2.14 \cdot 10^{-4}$ \\
\multirow[l]{2}{*}{Counterprop.} & BARQ & 3.35 & 3.34 & 2.08 & $9.74 \cdot 10^{-7}$ & $1.07 \cdot 10^{-6}$ & $4.35 \cdot 10^{-6}$ & $6.85 \cdot 10^{-4}$ & $6.18 \cdot 10^{-4}$ & $4.26 \cdot 10^{-4}$ \\
 & Constant & 0.19 & 0.16 & 0.24 & $1.47 \cdot 10^{-2}$ & $1.86 \cdot 10^{-2}$ & $7.54 \cdot 10^{-3}$ & $6.49 \cdot 10^{-4}$ & $3.72 \cdot 10^{-4}$ & $1.66 \cdot 10^{-4}$ \\
\end{tabular}

       \caption{Extended data for Table \ref{decay_times}.}
              \label{decay_times_extended}
    \end{ruledtabular}
\end{table*}

\begin{table*}
    \begin{ruledtabular}
 \setlength{\tabcolsep}{7pt}
 \begin{tabular}{llllllllllll}
 &  & \multicolumn{5}{c}{Diamond distance} & \multicolumn{5}{c}{Infidelity} \\
 & Qubit & q[0] & q[1] & q[2] & q[3] & Avg & q[0] & q[1] & q[2] & q[3] & Avg \\
Gate & BARQ Counterprop. over Any Coprop. &  &  &  &  &  &  &  &  &  &  \\
\multirow[l]{2}{*}{$X_{\pi/2}$} & BARQ & 0.61 & 0.37 & 0.31 & 0.58 & 0.48 & 1.47 & 0.89 & 0.69 & 1.15 & 1.04 \\
 & Constant & 0.83 & 0.30 & 0.44 & 1.56 & 0.68 & 1.04 & 0.54 & 0.68 & 1.28 & 0.86 \\
\multirow[l]{2}{*}{$Y_{\pi/2}$} & BARQ & 0.59 & 0.40 & 0.38 & 0.59 & 0.50 & 1.83 & 1.14 & 0.96 & 1.32 & 1.29 \\
 & Constant & 1.28 & 0.43 & 0.46 & 1.49 & 0.81 & 1.10 & 0.62 & 0.72 & 1.27 & 0.90 \\
\end{tabular}

\caption{Diamond distance and gate infidelity ratios between BARQ counterpropagating pulses and any type of copropagating pulses.}
\label{dd_infid_ratios}
    \end{ruledtabular}
\end{table*}

\begin{table*}
    \begin{ruledtabular}
 \setlength{\tabcolsep}{6pt}
 \begin{tabular}{cccccccccccc}
Qubit & q[0] & q[1] & q[2] & q[3] \\
RB fidelity & 99.82 $\pm$ $2.31 \cdot 10^{-2}$ \% & 99.87 $\pm$ $1.88 \cdot 10^{-2}$ \% & 99.77 $\pm$ $2.62 \cdot 10^{-2}$ \% & 99.78 $\pm$ $2.00 \cdot 10^{-2}$ \% \\
\end{tabular}

\caption{Randomized benchmarking experimental data for BARQ pulses in the counterpropagating beam configuration. We use pyGSTi to perform an RB experiment as done in  Ref.~\cite{WRIGHTBenchmarking11qubitQuantumComputer2019}. We choose 500 shots and RB lengths $L \in [2,4,6,8,10,12]$ and fit the resulting decay curve to $Bp^L + \frac{1}{2}$ where $p$ captures the average error in a gate-set circuit fidelity sense \cite{CARIGNAN-DUGASRandomizedBenchmarkingExperimentsGateset2018}. For each length $L$, we used 24 random Clifford sequences and compiled each Clifford based on the gate set in Sec. \ref{gst_results_sec}.}
\label{rb_data}
    \end{ruledtabular}
\end{table*}

\begin{table*}
    \begin{ruledtabular}
    \begin{tabular}{cccccccccc}
 &  &  &  & Diamond distance & Infidelity \\
Gate & Pulse & Beams & Qubit &  &  \\
\multirow[l]{20}{*}{$X_{\pi/2}$} & \multirow[l]{3}{*}{BARQ} & \multirow[l]{3}{*}{Counterprop.} & q[1] & $3.59 \cdot 10^{-3}$ $\pm$ $1.25 \cdot 10^{-3}$ & $1.36 \cdot 10^{-3}$ $\pm$ $9.04 \cdot 10^{-5}$ \\
 &  &  & q[2] & $3.84 \cdot 10^{-3}$ $\pm$ $1.15 \cdot 10^{-3}$ & $1.49 \cdot 10^{-3}$ $\pm$ $9.60 \cdot 10^{-5}$ \\
 &  &  & Avg & $6.15 \cdot 10^{-3}$ $\pm$ $1.41 \cdot 10^{-3}$ & $1.94 \cdot 10^{-3}$ $\pm$ $1.32 \cdot 10^{-4}$ \\
\cline{2-4} \cline{3-4}
 & Constant & Coprop. & q[3] & $6.26 \cdot 10^{-3}$ $\pm$ $2.50 \cdot 10^{-3}$ & $1.86 \cdot 10^{-3}$ $\pm$ $8.27 \cdot 10^{-5}$ \\
\cline{2-4} \cline{3-4}
 & BARQ & Counterprop. & q[0] & $7.43 \cdot 10^{-3}$ $\pm$ $1.93 \cdot 10^{-3}$ & $2.53 \cdot 10^{-3}$ $\pm$ $1.77 \cdot 10^{-4}$ \\
\cline{2-4} \cline{3-4}
 & \multirow[l]{4}{*}{Constant} & Counterprop. & q[3] & $8.47 \cdot 10^{-3}$ $\pm$ $4.60 \cdot 10^{-3}$ & $3.00 \cdot 10^{-3}$ $\pm$ $1.65 \cdot 10^{-4}$ \\
\cline{3-4}
 &  & \multirow[l]{3}{*}{Coprop.} & q[2] & $8.78 \cdot 10^{-3}$ $\pm$ $2.17 \cdot 10^{-3}$ & $2.19 \cdot 10^{-3}$ $\pm$ $1.40 \cdot 10^{-4}$ \\
 &  &  & Avg & $8.98 \cdot 10^{-3}$ $\pm$ $2.66 \cdot 10^{-3}$ & $2.25 \cdot 10^{-3}$ $\pm$ $1.14 \cdot 10^{-4}$ \\
 &  &  & q[0] & $9.00 \cdot 10^{-3}$ $\pm$ $2.39 \cdot 10^{-3}$ & $2.44 \cdot 10^{-3}$ $\pm$ $1.06 \cdot 10^{-4}$ \\
\cline{2-4} \cline{3-4}
 & \multirow[l]{2}{*}{BARQ} & Coprop. & q[1] & $9.69 \cdot 10^{-3}$ $\pm$ $2.01 \cdot 10^{-3}$ & $1.52 \cdot 10^{-3}$ $\pm$ $1.07 \cdot 10^{-4}$ \\
\cline{3-4}
 &  & Counterprop. & q[3] & $9.75 \cdot 10^{-3}$ $\pm$ $1.30 \cdot 10^{-3}$ & $2.38 \cdot 10^{-3}$ $\pm$ $1.63 \cdot 10^{-4}$ \\
\cline{2-4} \cline{3-4}
 & \multirow[l]{2}{*}{Constant} & Counterprop. & q[1] & $9.81 \cdot 10^{-3}$ $\pm$ $1.18 \cdot 10^{-3}$ & $2.28 \cdot 10^{-3}$ $\pm$ $1.28 \cdot 10^{-4}$ \\
\cline{3-4}
 &  & Coprop. & q[1] & $1.19 \cdot 10^{-2}$ $\pm$ $3.60 \cdot 10^{-3}$ & $2.50 \cdot 10^{-3}$ $\pm$ $1.29 \cdot 10^{-4}$ \\
\cline{2-4} \cline{3-4}
 & \multirow[l]{3}{*}{BARQ} & \multirow[l]{3}{*}{Coprop.} & q[2] & $1.22 \cdot 10^{-2}$ $\pm$ $2.73 \cdot 10^{-3}$ & $2.16 \cdot 10^{-3}$ $\pm$ $1.50 \cdot 10^{-4}$ \\
 &  &  & q[0] & $1.22 \cdot 10^{-2}$ $\pm$ $2.50 \cdot 10^{-3}$ & $1.72 \cdot 10^{-3}$ $\pm$ $1.12 \cdot 10^{-4}$ \\
 &  &  & Avg & $1.27 \cdot 10^{-2}$ $\pm$ $2.15 \cdot 10^{-3}$ & $1.87 \cdot 10^{-3}$ $\pm$ $1.25 \cdot 10^{-4}$ \\
\cline{2-4} \cline{3-4}
 & \multirow[l]{2}{*}{Constant} & \multirow[l]{2}{*}{Counterprop.} & Avg & $1.38 \cdot 10^{-2}$ $\pm$ $2.54 \cdot 10^{-3}$ & $2.80 \cdot 10^{-3}$ $\pm$ $1.49 \cdot 10^{-4}$ \\
 &  &  & q[2] & $1.61 \cdot 10^{-2}$ $\pm$ $1.43 \cdot 10^{-3}$ & $2.41 \cdot 10^{-3}$ $\pm$ $1.21 \cdot 10^{-4}$ \\
\cline{2-4} \cline{3-4}
 & BARQ & Coprop. & q[3] & $1.69 \cdot 10^{-2}$ $\pm$ $1.35 \cdot 10^{-3}$ & $2.07 \cdot 10^{-3}$ $\pm$ $1.30 \cdot 10^{-4}$ \\
\cline{2-4} \cline{3-4}
 & Constant & Counterprop. & q[0] & $2.10 \cdot 10^{-2}$ $\pm$ $2.94 \cdot 10^{-3}$ & $3.49 \cdot 10^{-3}$ $\pm$ $1.82 \cdot 10^{-4}$ \\
\cline{1-4} \cline{2-4} \cline{3-4}
\multirow[l]{20}{*}{$Y_{\pi/2}$} & \multirow[l]{2}{*}{BARQ} & \multirow[l]{2}{*}{Counterprop.} & q[1] & $3.80 \cdot 10^{-3}$ $\pm$ $1.33 \cdot 10^{-3}$ & $1.51 \cdot 10^{-3}$ $\pm$ $1.02 \cdot 10^{-4}$ \\
 &  &  & q[2] & $4.42 \cdot 10^{-3}$ $\pm$ $1.69 \cdot 10^{-3}$ & $1.75 \cdot 10^{-3}$ $\pm$ $1.22 \cdot 10^{-4}$ \\
\cline{2-4} \cline{3-4}
 & Constant & Coprop. & q[0] & $5.35 \cdot 10^{-3}$ $\pm$ $2.03 \cdot 10^{-3}$ & $2.46 \cdot 10^{-3}$ $\pm$ $1.11 \cdot 10^{-4}$ \\
\cline{2-4} \cline{3-4}
 & BARQ & Counterprop. & Avg & $6.01 \cdot 10^{-3}$ $\pm$ $1.46 \cdot 10^{-3}$ & $2.05 \cdot 10^{-3}$ $\pm$ $1.43 \cdot 10^{-4}$ \\
\cline{2-4} \cline{3-4}
 & \multirow[l]{2}{*}{Constant} & Coprop. & q[3] & $6.02 \cdot 10^{-3}$ $\pm$ $2.43 \cdot 10^{-3}$ & $1.77 \cdot 10^{-3}$ $\pm$ $9.96 \cdot 10^{-5}$ \\
\cline{3-4}
 &  & Counterprop. & q[3] & $6.45 \cdot 10^{-3}$ $\pm$ $3.51 \cdot 10^{-3}$ & $3.13 \cdot 10^{-3}$ $\pm$ $1.69 \cdot 10^{-4}$ \\
\cline{2-4} \cline{3-4}
 & BARQ & Counterprop. & q[0] & $6.84 \cdot 10^{-3}$ $\pm$ $1.45 \cdot 10^{-3}$ & $2.69 \cdot 10^{-3}$ $\pm$ $1.93 \cdot 10^{-4}$ \\
\cline{2-4} \cline{3-4}
 & \multirow[l]{3}{*}{Constant} & Coprop. & Avg & $7.45 \cdot 10^{-3}$ $\pm$ $2.74 \cdot 10^{-3}$ & $2.27 \cdot 10^{-3}$ $\pm$ $1.14 \cdot 10^{-4}$ \\
\cline{3-4}
 &  & Counterprop. & q[1] & $8.43 \cdot 10^{-3}$ $\pm$ $4.02 \cdot 10^{-3}$ & $2.30 \cdot 10^{-3}$ $\pm$ $1.31 \cdot 10^{-4}$ \\
\cline{3-4}
 &  & Coprop. & q[1] & $8.85 \cdot 10^{-3}$ $\pm$ $4.46 \cdot 10^{-3}$ & $2.45 \cdot 10^{-3}$ $\pm$ $1.24 \cdot 10^{-4}$ \\
\cline{2-4} \cline{3-4}
 & \multirow[l]{2}{*}{BARQ} & Counterprop. & q[3] & $8.98 \cdot 10^{-3}$ $\pm$ $1.36 \cdot 10^{-3}$ & $2.24 \cdot 10^{-3}$ $\pm$ $1.56 \cdot 10^{-4}$ \\
\cline{3-4}
 &  & Coprop. & q[1] & $9.56 \cdot 10^{-3}$ $\pm$ $1.10 \cdot 10^{-3}$ & $1.32 \cdot 10^{-3}$ $\pm$ $7.79 \cdot 10^{-5}$ \\
\cline{2-4} \cline{3-4}
 & Constant & Coprop. & q[2] & $9.57 \cdot 10^{-3}$ $\pm$ $2.03 \cdot 10^{-3}$ & $2.42 \cdot 10^{-3}$ $\pm$ $1.23 \cdot 10^{-4}$ \\
\cline{2-4} \cline{3-4}
 & \multirow[l]{3}{*}{BARQ} & \multirow[l]{3}{*}{Coprop.} & q[0] & $1.16 \cdot 10^{-2}$ $\pm$ $3.03 \cdot 10^{-3}$ & $1.47 \cdot 10^{-3}$ $\pm$ $9.45 \cdot 10^{-5}$ \\
 &  &  & q[2] & $1.17 \cdot 10^{-2}$ $\pm$ $3.46 \cdot 10^{-3}$ & $1.83 \cdot 10^{-3}$ $\pm$ $1.47 \cdot 10^{-4}$ \\
 &  &  & Avg & $1.20 \cdot 10^{-2}$ $\pm$ $2.16 \cdot 10^{-3}$ & $1.58 \cdot 10^{-3}$ $\pm$ $9.22 \cdot 10^{-5}$ \\
\cline{2-4} \cline{3-4}
 & \multirow[l]{2}{*}{Constant} & \multirow[l]{2}{*}{Counterprop.} & Avg & $1.24 \cdot 10^{-2}$ $\pm$ $2.56 \cdot 10^{-3}$ & $2.82 \cdot 10^{-3}$ $\pm$ $1.56 \cdot 10^{-4}$ \\
 &  &  & q[2] & $1.41 \cdot 10^{-2}$ $\pm$ $1.17 \cdot 10^{-3}$ & $2.35 \cdot 10^{-3}$ $\pm$ $1.29 \cdot 10^{-4}$ \\
\cline{2-4} \cline{3-4}
 & BARQ & Coprop. & q[3] & $1.52 \cdot 10^{-2}$ $\pm$ $1.04 \cdot 10^{-3}$ & $1.70 \cdot 10^{-3}$ $\pm$ $4.92 \cdot 10^{-5}$ \\
\cline{2-4} \cline{3-4}
 & Constant & Counterprop. & q[0] & $2.08 \cdot 10^{-2}$ $\pm$ $1.54 \cdot 10^{-3}$ & $3.50 \cdot 10^{-3}$ $\pm$ $1.94 \cdot 10^{-4}$ \\
\cline{1-4} \cline{2-4} \cline{3-4}
\multirow[l]{19}{*}{$Z_{\pi/2}$} & \multirow[l]{5}{*}{Constant} & Coprop. & q[2] & $1.20 \cdot 10^{-3}$ $\pm$ $1.15 \cdot 10^{-3}$ & $2.61 \cdot 10^{-6}$ $\pm$ $1.27 \cdot 10^{-5}$ \\
\cline{3-4}
 &  & \multirow[l]{2}{*}{Counterprop.} & q[3] & $1.48 \cdot 10^{-3}$ $\pm$ $4.89 \cdot 10^{-4}$ & $2.25 \cdot 10^{-6}$ $\pm$ $1.46 \cdot 10^{-5}$ \\
 &  &  & q[2] & $2.04 \cdot 10^{-3}$ $\pm$ $1.12 \cdot 10^{-3}$ & $3.58 \cdot 10^{-6}$ $\pm$ $1.76 \cdot 10^{-5}$ \\
\cline{3-4}
 &  & Coprop. & q[3] & $2.07 \cdot 10^{-3}$ $\pm$ $1.99 \cdot 10^{-3}$ & $3.01 \cdot 10^{-6}$ $\pm$ $1.36 \cdot 10^{-5}$ \\
\cline{3-4}
 &  & Counterprop. & Avg & $2.42 \cdot 10^{-3}$ $\pm$ $9.86 \cdot 10^{-4}$ & $6.12 \cdot 10^{-6}$ $\pm$ $1.62 \cdot 10^{-5}$ \\
\cline{2-4} \cline{3-4}
 & BARQ & Counterprop. & q[1] & $2.68 \cdot 10^{-3}$ $\pm$ $1.44 \cdot 10^{-3}$ & $1.20 \cdot 10^{-5}$ $\pm$ $1.66 \cdot 10^{-5}$ \\
\cline{2-4} \cline{3-4}
 & \multirow[l]{5}{*}{Constant} & \multirow[l]{2}{*}{Counterprop.} & q[0] & $3.02 \cdot 10^{-3}$ $\pm$ $1.15 \cdot 10^{-3}$ & $6.75 \cdot 10^{-6}$ $\pm$ $1.64 \cdot 10^{-5}$ \\
 &  &  & q[1] & $3.14 \cdot 10^{-3}$ $\pm$ $1.19 \cdot 10^{-3}$ & $1.19 \cdot 10^{-5}$ $\pm$ $1.61 \cdot 10^{-5}$ \\
\cline{3-4}
 &  & \multirow[l]{3}{*}{Coprop.} & Avg & $3.37 \cdot 10^{-3}$ $\pm$ $3.16 \cdot 10^{-3}$ & $1.22 \cdot 10^{-5}$ $\pm$ $2.06 \cdot 10^{-5}$ \\
 &  &  & q[1] & $4.60 \cdot 10^{-3}$ $\pm$ $6.05 \cdot 10^{-3}$ & $2.20 \cdot 10^{-5}$ $\pm$ $3.98 \cdot 10^{-5}$ \\
 &  &  & q[0] & $5.64 \cdot 10^{-3}$ $\pm$ $3.45 \cdot 10^{-3}$ & $2.12 \cdot 10^{-5}$ $\pm$ $1.64 \cdot 10^{-5}$ \\
\cline{2-4} \cline{3-4}
 & \multirow[l]{8}{*}{BARQ} & \multirow[l]{3}{*}{Counterprop.} & Avg & $5.90 \cdot 10^{-3}$ $\pm$ $1.41 \cdot 10^{-3}$ & $3.75 \cdot 10^{-5}$ $\pm$ $2.00 \cdot 10^{-5}$ \\
 &  &  & q[0] & $6.03 \cdot 10^{-3}$ $\pm$ $1.56 \cdot 10^{-3}$ & $3.75 \cdot 10^{-5}$ $\pm$ $1.92 \cdot 10^{-5}$ \\
 &  &  & q[3] & $8.98 \cdot 10^{-3}$ $\pm$ $1.25 \cdot 10^{-3}$ & $6.31 \cdot 10^{-5}$ $\pm$ $2.42 \cdot 10^{-5}$ \\
\cline{3-4}
 &  & \multirow[l]{5}{*}{Coprop.} & q[1] & $1.00 \cdot 10^{-2}$ $\pm$ $2.84 \cdot 10^{-3}$ & $6.77 \cdot 10^{-5}$ $\pm$ $1.37 \cdot 10^{-5}$ \\
 &  &  & q[0] & $1.41 \cdot 10^{-2}$ $\pm$ $7.46 \cdot 10^{-4}$ & $1.32 \cdot 10^{-4}$ $\pm$ $1.69 \cdot 10^{-5}$ \\
 &  &  & q[2] & $1.41 \cdot 10^{-2}$ $\pm$ $6.43 \cdot 10^{-3}$ & $1.37 \cdot 10^{-4}$ $\pm$ $7.13 \cdot 10^{-5}$ \\
 &  &  & Avg & $1.45 \cdot 10^{-2}$ $\pm$ $3.21 \cdot 10^{-3}$ & $1.50 \cdot 10^{-4}$ $\pm$ $3.39 \cdot 10^{-5}$ \\
 &  &  & q[3] & $1.99 \cdot 10^{-2}$ $\pm$ $2.80 \cdot 10^{-3}$ & $2.64 \cdot 10^{-4}$ $\pm$ $3.37 \cdot 10^{-5}$ \\
\cline{1-4} \cline{2-4} \cline{3-4}
\end{tabular}

        \end{ruledtabular}
    \caption{Diamond distance and gate infidelity data used for Fig. \ref{gst_results}, sorted by diamond distance values.}
    \label{gst_results_table}
\end{table*}

\end{document}